# About the size of Google Scholar: playing the numbers

**Enrique Orduña-Malea[1], Juan Manuel Ayllón[2], Alberto Martín-Martín[2], Emilio Delgado López-Cózar[2]**

[1] *EC3: Evaluación de la Ciencia y de la Comunicación Científica, Universidad Politécnica de Valencia (Spain)*
[2] *EC3: Evaluación de la Ciencia y de la Comunicación Científica, Universidad de Granada (Spain)*

## ABSTRACT

The emergence of academic search engines (Google Scholar and Microsoft Academic Search essentially) has revived and increased the interest in the size of the academic web, since their aspiration is to index the entirety of current academic knowledge. The search engine functionality and human search patterns lead us to believe, sometimes, that what you see in the search engine's results page is all that really exists. And, even when this is not true, we wonder which information is missing and why. The main objective of this working paper is to calculate the size of Google Scholar at present (May 2014). To do this, we present, apply and discuss up to 4 empirical methods: Khabsa & Giles's method, an estimate based on empirical data, and estimates based on direct queries and absurd queries. The results, despite providing disparate values, place the estimated size of Google Scholar in about 160 million documents. However, the fact that all methods show great inconsistencies, limitations and uncertainties, makes us wonder why Google does not simply provide this information to the scientific community if the company really knows this figure.

**Corresponding author**
**Emilio Delgado López-Cózar.** edelgado@ugr.es
**Enrique Orduña-Malea.** enorma@upv.es

# 1. INTRODUCTION

The calculation of the size of the Web in general (Lawrence & Giles, 1998; 1999; Dobra & Fienberg, 2004, among others) and the academic web in particular (Khabsa & Giles, 2014) has generated a debate in the scientific arena over the last two decades at different levels, among which we can highlight the following: a) from an information perspective (the extent to which all the knowledge produced is actually indexed, searchable, retrievable and accessible from a catalogue, index or database); b) from a methodological level (how to calculate it as accurately as possible); and c) from a socioeconomic perspective (how the composition and evolution of these contents affect their consumption in different countries according to different social, economic and political issues).

The emergence of academic search engines (Ortega, 2014), Google Scholar and Microsoft Academic Search essentially, has revived and increased the interest in the size of the academic web, changing the focus of the question: their aspiration to index the entirety of current academic knowledge leads us to believe, sometimes, that what you see in the search engine result page is all that really exists. And, even when this is not true, we wonder which information is missing.

In traditional bibliographic databases (WoS, Scopus), finding out the size (measured by the number of records at a given time) is a fairly trivial matter (it's only necessary to perform a query in the search interface), because the entire universe is catalogued and under control (always accounting for a low error rate due to a lack of absolute normalization in the catalogue). Moreover, the evolution of these databases is cumulative: the number of records always grows and never decreases, except for the exceptional elimination of records due to technical or legal issues. However, in the case of academic search engines, these assertions not always apply, making both the calculation of their size, and tracking the evolution of their data, extremely complicated tasks.

Despite the high dynamism of the Web (contents are continually added, changed and/or deleted worldwide) and the inherent technical difficulties to catalogue and update such a vast and diverse universe (Koehler, 2004), the main problem is the opaque information policies followed by those responsible for such databases, particularly Google Scholar (GS). Its policies differ from those of other discovery services like Microsoft Academic Search (MAS) or Web of Science (WoS), where just running a query that aims to get the number of documents indexed in a specific date will get you the answer instantly.

The recent work of Khabsa and Giles (2014) has estimated the number of circulating documents written in English in the academic Web on 114 million (and also that of those, GS has around 99.8 million), employing a procedure based on the use of the Lincolm-Petersen (capture-recapture) method, from the citations to a sample of articles written in English, included both in GS and MAS. However, this procedure (discussed in greater detail in the methods and results sections) leads us to formulate a number of questions, namely: is it possible to calculate the size of the academic web in general (and Google

Scholar in particular)? And, does Google Scholar really cover 87% of the global academic Web?

## 2. OBJECTIVES

The main objective of this working paper is to calculate the size of Google Scholar at present (May 2014). To do this, the specific objectives are the following:

- Explain and apply various empirical methods to estimate the size of Google Scholar.
- Point up the strengths and weaknesses of the different methods for estimating the size of Google Scholar.

## 3. METHODS

To estimate the size of Google Scholar we propose and test 4 different procedures, explained in further detail below:

### a) Khabsa & Giles's method

This procedure is taken directly from the research carried out by Khabsa and Giles (2014), and recently "digested" by Orduña-Malea et al (2014). The method is as follows:

First, 150 English academic papers (journal and conference papers, dissertations and masters theses, books, technical reports and working papers) are selected from Microsoft Academic Search (MAS). These articles are randomly sampled from the most cited documents in each of the 15 fields covered by MAS (10 documents per field), considering only documents with less than 1,000 citations. These 150 articles are verified to be included in Google Scholar as well.

After this, the number of incoming citations to the 150 selected documents are obtained both from MAS (41,778 citations) and Google Scholar (86,870). The overlap between GS and MAS (citing documents contained in both search engines) is computed by means of the Jaccard similarity index (0.418). These data is collected in January 2013.

The number of scholarly documents available on the web is estimated then using the Lincoln-Petersen method (capture/recapture):

$$\frac{R}{M} = \frac{C}{N}$$

Where**:**

**N** (population) = size of GS + size of MAS;
**M** (elements captured in the first sample) = size of GS;
**C** (elements captured in the second sample) = size of MAS;

**R** (elements recaptured in the second sample) = overlap between GS and MAS, measured as the number of citations shared).

Note that the expression on the left side of the equal symbol is the original meaning of the Lincolm-Petersen indicators, and the one on the right side is the analogy used by Khabsa and Giles. It is also noteworthy that whereas M and C correspond to the number of documents indexed on GS and MAS respectively, the overlap between them (R) is measured through the citing documents to the cited documents of the sample.

Since C is taken from the data provided by Microsoft (48,774,764 million documents, which is reduced to 47,799,627 after applying a correction factor for English of 0.98), and R is calculated directly by the authors, then N can be directly isolated, and subsequently M (the size of Google Scholar).

**b) Estimates from empirical data**

The second method consists of making estimates from empirical studies that have previously worked with samples and have compared GS with other databases. From these comparisons and the differences in coverage, a correction factor could be obtained, and consequently a hypothetical projection may be proposed.

To this end, an extensive collection of empirical studies dealing with the calculation of the sizes of academic databases has been gathered in Appendix I. It shows, in table format, each collected work, the database analysed (GS, WoS, MAS, Scopus, Pubmed, etc.), the unit of analysis (citations, documents, etc.) and the sample considered.

These studies use different units of analysis (journals, articles, books, etc.) and metrics (citation count, h-index, impact factor, etc.), but for our purpose, the synthesis of the results can only be applied to samples that are comparable to each other in the following levels:

- Studies examining the same databases used as data source.
- Studies working with documents or with unique citation documents.
- Studies that make comparisons between documents written in the same language (or do not make a distinction by language).

Hence, we have categorized the data offered in Appendix I according to the unit of study: journals, books, etc.; the indicator measured: citations, documents, citations per document (Appendix II); and according to the language of the documents (Appendix III).

Finally, only those studies comparing GS and WoS have been considered, since only just two studies provided information about empirical comparisons between GS and Scopus, and the remaining databases were even less well represented.

Then, for each case study we obtained the proportion between both databases dividing the number of documents retrieved for GS by the number of documents gathered for WoS. Finally, the geometric mean and the median of all studies are carried out to get a crude, but indicative, correction factor.

This same procedure has also been applied to the comparison of unique citing documents as unit of analysis (citing documents indexed in GS and non-indexed in WoS, and vice versa).

**c) Direct query**

The third method is based on interrogating the database itself, at least to the extent that this is possible. This can be done by two procedures: a) using the custom date range for the complete period of time; b) using the custom date range year by year and adding the results together at the end.

To this end, first we directly queried (by means of an empty query search) Google Scholar (the <google.scholar.com> version) filtering by single years[1] and gathering the estimated number of results, also called Hit Count Estimates (HCE). We have processed the data from 1700 to 2013 (data prior to 1700 are practically non-existent; namely, 49 records are found in the range 1000-1700). After this, the number of records obtained for each year are added together.

At the same time, we set the custom range from 1700 to 2013, to gather all documents in the period in the same query. This data collection process was carried out in May 2014.

The raw hit count estimates is comprised of three different types of results (some of which may be included or excluded in a query): ordinary records (documents indexed on Google Scholar, providing a link to the full text or to a paid gateway), citations (references to documents not indexed on Google Scholar), and patents (documents extracted from Google Patents).

To test the potential influence of citations and patents in the size of Google Scholar, we retrieved the following data for each year:

- All documents (records + citations + patents);
- Records + citations;
- Records + patents;
- Only records, excluding citations and patents.

Finally, direct queries were performed on Microsoft Academic Search as well as on Web of Science Core Collection , with the aim of gathering both sectional and longitudinal data about the size of these databases at present (data were collected in May 2014) suitable to compare with those obtained previously for Google Scholar:

- Microsoft Academic Search: from <academic.research.microsoft.com>, direct query via the "year" command is performed.

- Web of Science Core Collection: the size is obtained from the basic search interface, specifying the year in the "Year published" field. Data concerning language and type of document were gathered as well.

For each database, a query per year (from 1700 to 2013) and a global query from 1700 were performed. In Figure 1 we show the interface of each of the 3 databases queried.

***Figure 1. Direct query performed against Microsoft Academic Search (top), Web of Science Core Collection (middle) and Google Scholar (bottom)***

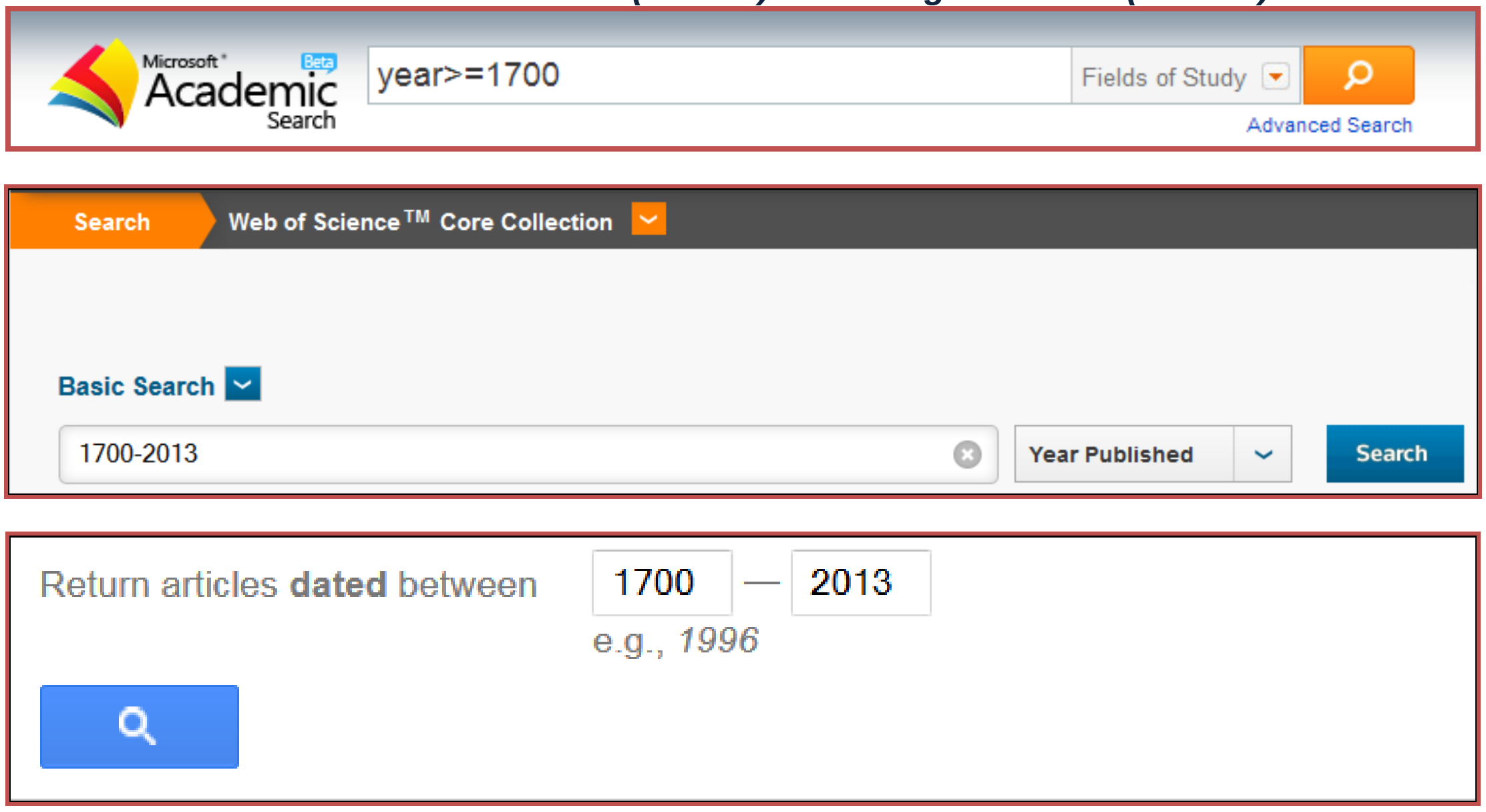

## d) Absurd query

The last method proposed is based on the use of some characteristics of the Boolean logic that are supported in Google Scholar's search box. In this case, the goal is to compose a query that somehow requires Google Scholar to return all its records. Although nowhere in the official documentation is it stated that such a query exists, we have run test queries using the following syntax: <common_term -site:unexistent_site>

The idea behind this is to query the occurrences of a very common term (likely to appear in almost all written records), and to filter out its appearances in a nonexistent site, which means that we are implicitly selecting every existing site in our query.

For example: <a -site:ssstfsffsdfasdfsf.com>, or <1 -site:ssstfsffsdfasdfsf.com>.

The reason for including a term before the "-site" command is that this command does not work on its own. These queries were run on Google Scholar (including and excluding citations and patents) in June 2014.

As in the case of the direct query method, the queries were performed in two different ways: a) setting the custom range from 1700 to 2013; b) running a query for each year and adding the results together.

# 4. RESULTS

Various estimates of the size of Google Scholar, as calculated from each of the 4 procedures outlined above, are offered, additionally providing a discussion about their advantages, disadvantages and shortcomings.

## 4.1. Khabsa & Giles's method

Khabsa and Giles (2014) estimate on 114 million the number of circulating documents written in English in the academic Web, and on 99.3 million the number of English documents in Google Scholar.

Nonetheless, these figures may be biased due to the following considerations: applying the Lincolm-Petersen method; considering GS and MAS as the whole academic web universe; collecting biased data; and considering a biased database for the estimation. Let's discuss each of these issues in greater detail.

### a) Lincolm-Petersen estimate

The problem is not only related to the possible growth of the population among samples, or the condition on the equal probability of each element to be recaptured (conditions in the application of this estimate method), but also to the assumption that each sample is applied to different universes (Google Scholar and Microsoft Academic Search), when the original method consists of 2 (or more) captures in the same universe.

The authors use a complementary method obtaining similar results, and this reinforces the results. In any case, a reasonable uncertainty exists that should at least be noted.

### b) Size of the Academic Web (N)

The estimate of the total size of the academic web was probably undersized: the summation of Google Scholar and Microsoft Academic Search is still far from representing the total academic web space, though it undoubtedly makes up a very high percentage.

N (i.e., the scholarly academic public web) is considered to be the summation of Google Scholar and Microsoft Academic Search. This issue keeps out other databases, such as Google Books (it is well-known that Google Scholar and Google Books databases do not entirely match), among others, although we are aware that these missing results are probably low and statistically insignificant.

Besides this, there is a more fundamental concern: the low indexation of institutional repositories on Google Scholar (Orduña-Malea and Delgado López-Cózar, in press) as well as some of GS's indexing policies (for example, files over 5MB are not indexed, a procedure which is especially critical for doctoral theses).

For these reasons, the assumption that Google Scholar covers 87% of the global academic Web (even considering only English documents) may constitute an over-representation.

Moreover, the method considers Microsoft Academic Search as a valid universe (48.7 million of documents, as of January 2013). However, the information about the total size of MAS is confusing at present. Microsoft Azure Marketplace shows (as of May 2014) 39.85 million documents, which does not match the data used in Khabsa & Giles's research; from the information collected on the Web, we can estimate 45.3 million documents, and 45.9 million documents if a query is performed manually in the website platform (as of May 2014). How can this disparate information affect the calculation of "N"?

**c) Biased sample**

On one hand, the sample of cited documents is not absolutely random because only documents in English with less than 1,000 citations are considered. The authors acknowledge this limitation: search engines impose a restriction on the number of retrievable results for all type of queries, unless an Application Programmable Interface (API) is provided (and Google Scholar does not provide an API at the moment).

Accessing to only the first 1,000 documents may bias the sample in an unknown way (and maybe differently for each field), although we can assume (though not demonstrate) that these records contain the more formalized, visible, and more circulating and cited documents. Moreover, this statistical error is equally distributed to the 15 samples, thus reducing its effect.

On the other hand, the sample is uniform for each field (10 articles) despite the fact that the output size of each field is quite different. This may introduce an important bias in the estimation.

**d) Biased database**

The data sample is taken from MAS, and this database is biased, among other reasons, due to the diversity of language and document types:

- It is oriented to collect English written literature, and specifically that produced in English-speaking countries.
- It is oriented to very specific types of publication: articles and conference papers. Although it has recently incorporated monographs, they are still a genuine minority. In contrast, document types in GS are not as skewed towards journal articles and conference as they are in MAS.

This concern applies not only to MAS but also to Scopus and WoS. Since it is not possible to take data from MAS according to language and document type, let us discuss these topics considering WoS as a basis for our examples.

*Biases in the sample by language*

The WoS database is clearly biased towards English, as is well known. In Figure 2 you can empirically test the proportion of languages for documents indexed in the Web of Science for the period 1900 to 2014, where English amounts to slightly over 90% of the records.

***Figure 2. Language of documents indexed in the Web of Science (1900-2014)***

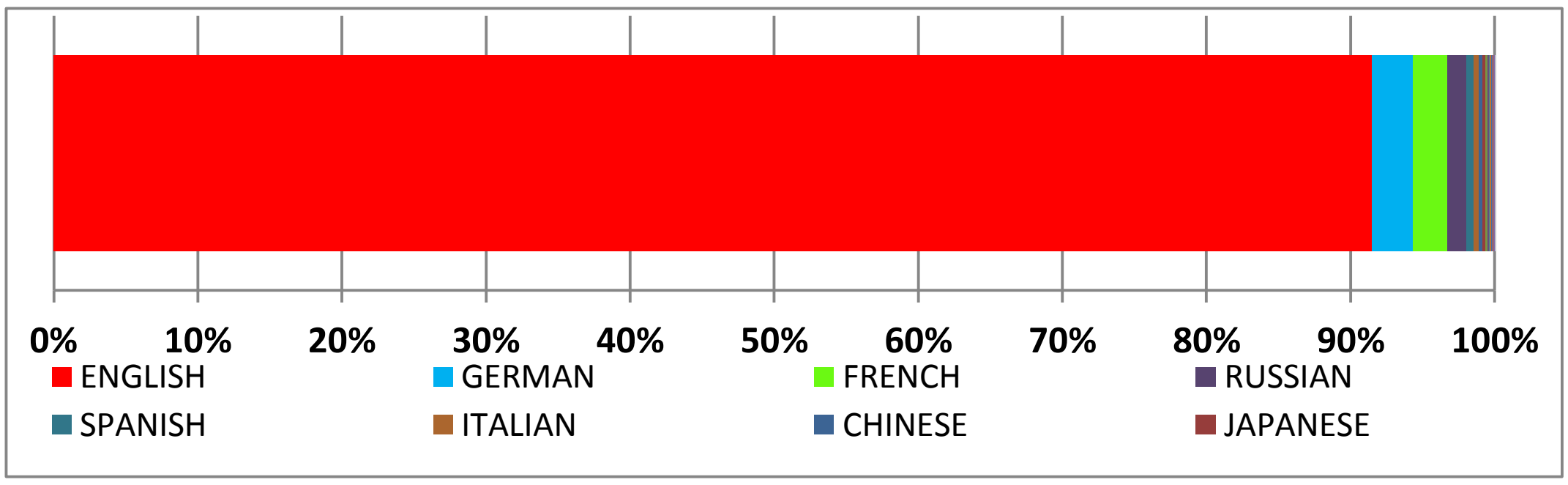

*Source: self-elaborated*

The proportion of English documents on MAS is high as well. Khabsa and Giles estimate in their sample a 98% of documents in English. However, the proportion of English documents on Google Scholar, though high, is lower than the one obtained in WoS and MAS. This different proportion may influence the estimate.

Otherwise, the estimated size of Google Scholar in English offered by Khabsa and Giles was probably oversized because, although the sample items (cited documents) were written in English, the citing documents could potentially be articles written in other languages.  This means that it is highly probable that the measurement was not entirely limited to the English universe.

*Biases in the sample by document type*

On the other hand, there is a bias towards articles. In Figure 3 we show the percentages of documents by type, collected in the Web of Science, for the period 1900 to 2014, where the “Journal document type” (composed by articles, meeting abstracts, editorial material and letters) represents 75% of all documents, whereas “Book and Book chapters” only 1%.

This bias creates a clear infra-representation of disciplines using other communication vehicles different than the "Journal article" format. Therefore, and as happened in the case of languages, the distribution of document types is not the same in WoS than in Google Scholar or Microsoft Academic Search.

*Figure 3. Document types indexed in the Web of Science (1900-2014)*

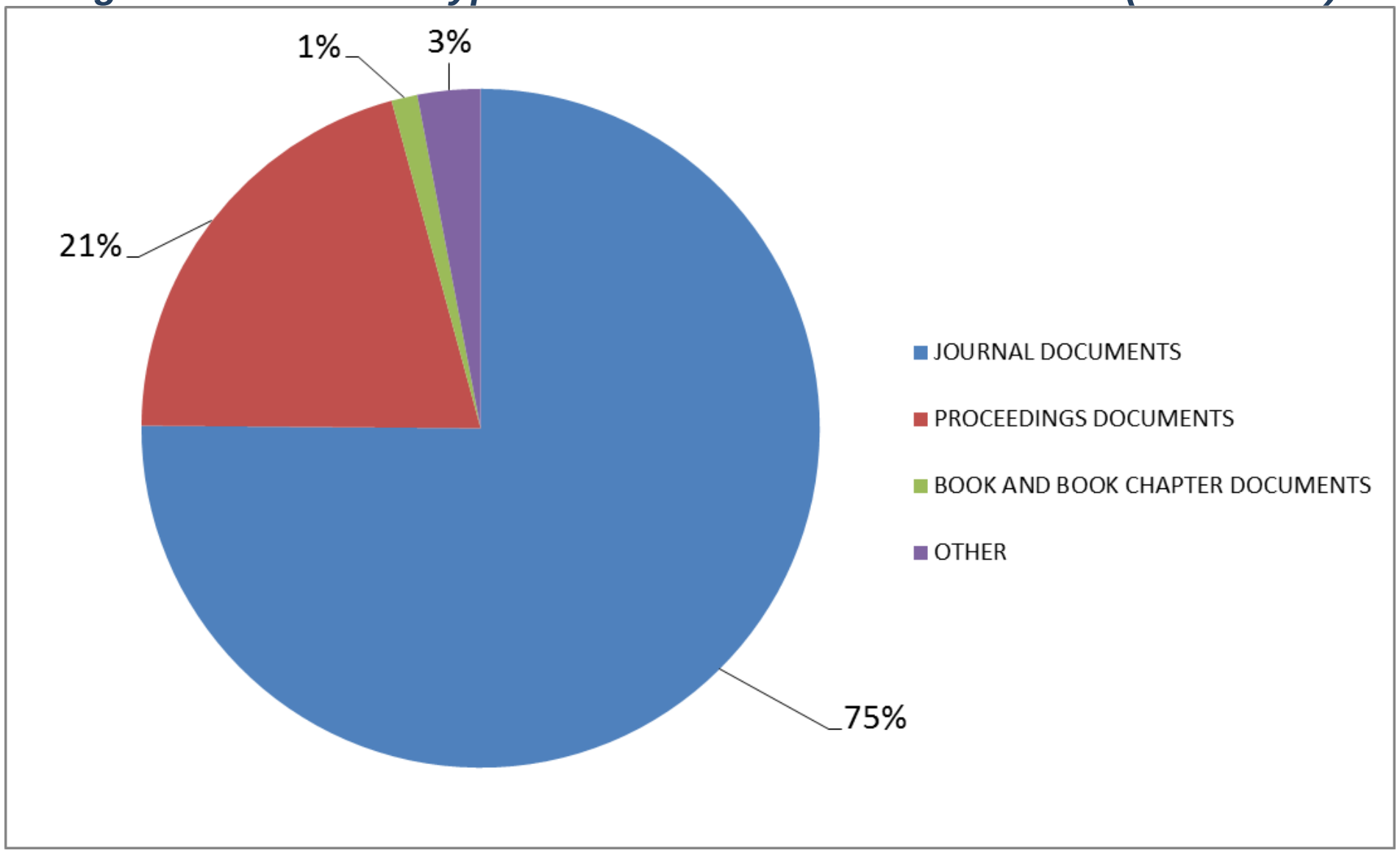

*Source: self-elaborated*

Unfortunately, you cannot perform this typological analysis directly on Google Scholar, because searching by type of document is not supported. Taking a look at the empirical data available in Appendix II, and considering that it is difficult to summarize all this information accurately because each study has its own distinct nature and deals with a different field, we can surmise that journal articles make up, on average, 65% of the total number of records in GS. In the case of Microsoft Academic Search, you can only manually check the number of document in articles and conferences.

If we carry out a global search in WorldCat[2] (the largest bibliographic information system in the world), approximately 2 billion items in more than 470 languages are obtained, but obviously this catalog covers both scientific and non-scientific (or digitized) documents. In any case, this procedure can itself help us to determine the proportion of books and other documents that are not included in traditional bibliographic databases, and which may not be indexed on Google Scholar as well. For example, a search for “thesis” (doctoral, masters or degree) gives a current figure of 16.3 million documents (Figure 4).

*Figure 4. Searching for Theses in Worldcat*

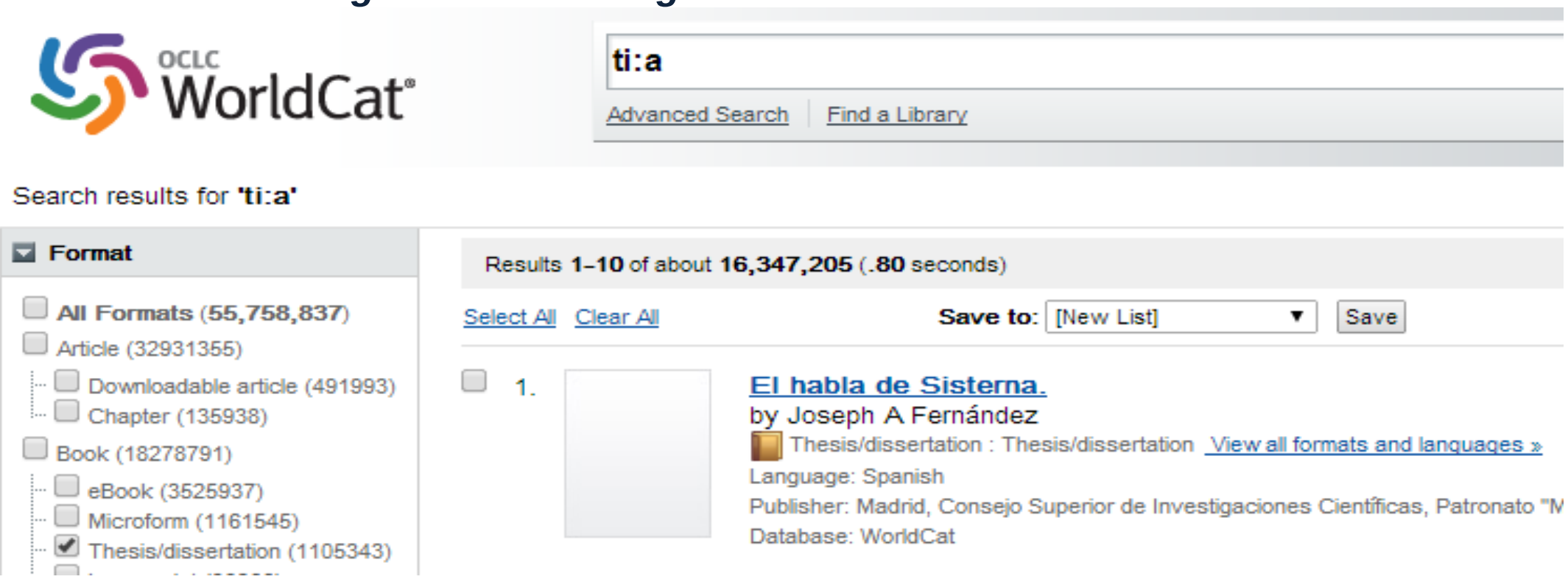

*Source: Worldcat*

Although Google Scholar indexes doctoral theses (since university institutional repositories are indexed, as well as some library services), this issue raises the following question: how many of these doctoral theses are indexed on Google Scholar, keeping in mind the limitation of a maximum of 5MB per file?, and how many are indexed on Microsoft Academic Search?

Moreover, if we analyze the production of scientific institutions, especially universities (the main producers of academic output), we can observe this production greatly varies from one university to another. Figure 5 offers the percentage of documents according to document type for various Spanish universities, such as the Polytechnic University of Catalonia (UPC), University of Barcelona (UB), Carlos III University of Madrid (UC3), and the aggregate value for all the universities in the region of Andalusia (collected from the Scientific information Service of Andalusia: SICA), as well as the overall average values. These empirical data can be extended by consulting the studies of Solis Cabrera (2008), Filippo et al (2011), the Annual Academic Report of the Complutense University of Madrid (*Vicerrectorado…, 2012*), and the FUTUR[3] web portal.

The data shown in Figure 5 indicates that journal articles, in general, are the most abundant type of publication, but that it still does not exceed 40% of the total production; books and book chapters amount to 30%, and conference communications come to around 20%.

***Figure 5 Scientific output according to document types in Spanish universities***

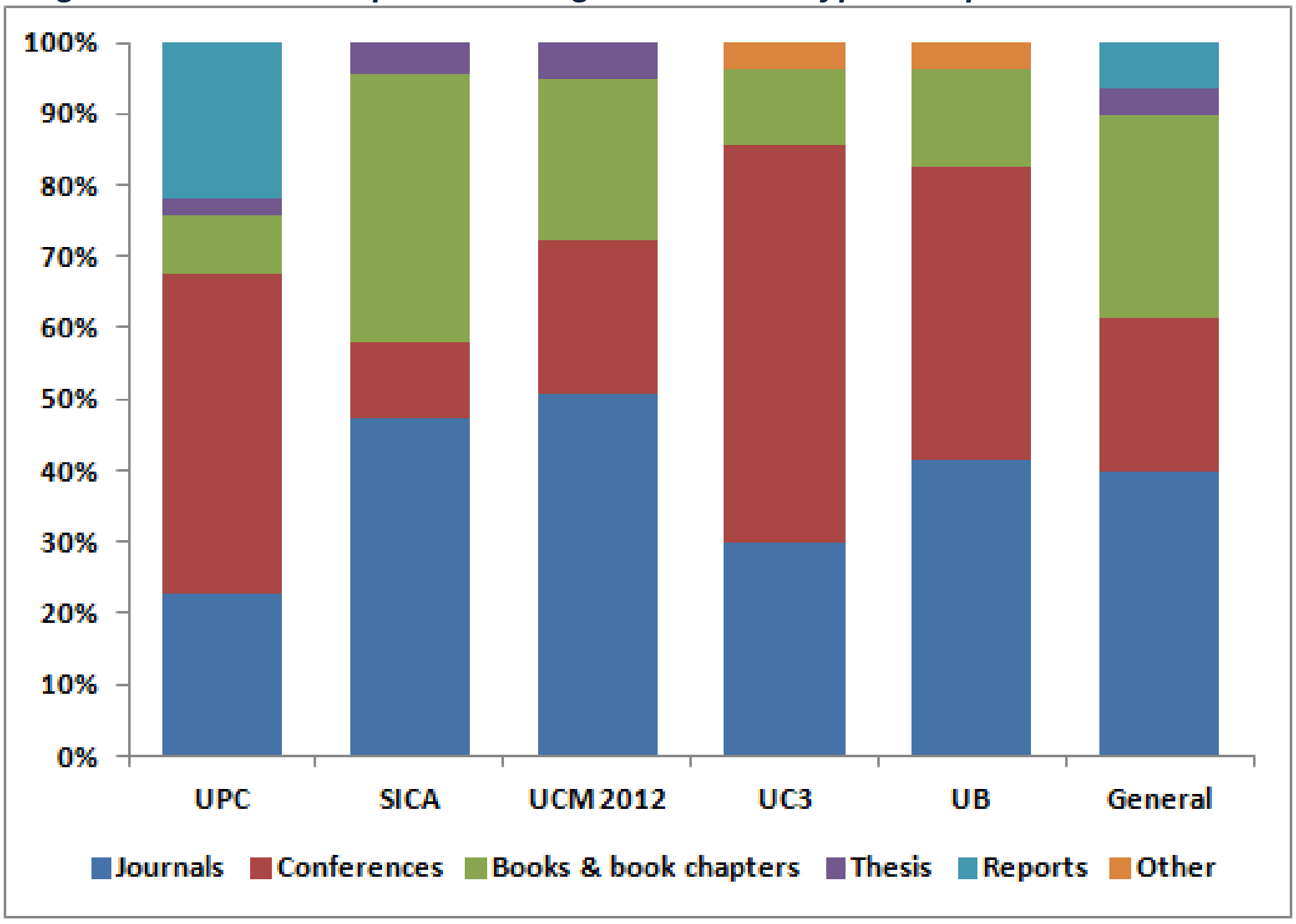

*Source: re-elaborated*

Clearly, databases such as WoS (or MAS and Scopus), heavily skewed towards the journal article format, cannot be used exclusively to estimate the size of the academic Web by means of the Lincolm-Petersen method, since they don't cover even 30% of the total scientific output of an institution, at least in the Spanish case (note that most of the journals or conferences where Spanish authors publish are not included in any of the aforementioned databases). Surely this same phenomenon occurs in other countries in a similar manner (always accounting for differences between university profiles).

Thus, the inferences from these databases, which are not representative of the entire collection of scientific literature (they are biased both by language and by document types), induce to not adequately estimate neither the size of the current academic literature in general nor the size of Google Scholar in particular.

### 4.2. Estimates from empirical data

In Table 1 we can observe the median, the geometric mean, and the number of studies that conforms the empirical set for each unit of study (number of documents and unique citing documents). The complete results obtained from the empirical data are available in Appendix I.

***Table 1. Correction factor obtained from empirical studies***

| UNIT OF STUDY | MEDIAN | GEOMETRIC MEAN | N* |
|---|---|---|---|
| Number of documents | 3 | 2.8 | 8 |
| Unique citing documents | 2.4 | 2.9 | 9 |

* The studies with less than 10 documents in the sample of WoS have not been finally considered since they are not representative enough.

We might assume (based on both empirical studies referenced in Appendix I and data provided in Table1) than at a general level (a round correction factor of 3), GS could triple the contents of WoS, although both databases would have a different English content distribution, and a different document type distribution. This is of particular importance as it is implying that the size comparison is not influenced by the biases of the database shown in the previous section.

Precisely, from the empirical studies, we test that the proportion of English documents in GS is around 65% (Appendix III). This would mean that, for documents in English, GS does not triple the number of WoS documents, but it probably does for documents in other languages. Knowing the general size correction factor, we don't need to worry about languages distribution for calculating estimates.

Therefore, the process of making inferences from samples of empirical studies previously conducted leads us in a simple way to multiply the size of WoS three times. As WoS has currently about 57 million records (see Figure 6), we may consider a quantity around 171 million records for GS.

These data and relationships can be expressed formally as follows (Table 2):

***Table 2. Size relationships between Web of Science (WoS) and Google Scholar (GS)***

| EQUATION | OBSERVATIONS |
| --- | --- |
| 3 * WoS = GS [1] | We apply a correction factor of 3: GS triples WoS. |
| WoS = WoSe*0.9 + WoSo*0.1 [2] | WoSe: English contents in WoS<br>WoSo: WoS content from other languages |
| GS = GSe*0.65 + GSo*0.35 [3] | GSe: English contents in GS (approx. 65%)<br>GSo: GS content from other languages |
| 3 * (WoSe*0.9+WoSo*0.1) = GSe*0.65 + GSo*0.35 | Substituting [2] and [3] in [1] |
| WoS = 57 million documents;<br>WoSe = 51.3 million documents; | We assume that WoS currently contains approximately 57 million records |

*Source: self-elaborated*

With these data in hand, we obtain about 111.15 million documents in English, and applying the 65% of English documents in the estimate by Khabsa & Giles (99.8 million), we get a total of 153.5 million documents. This is an unexpected result because, as we discussed previously, we previously thought that Khabsa & Giles's method was overestimating the number of English documents in Google Scholar.

The estimates from empirical data, however, present some important shortcomings as well, since it is difficult to synthesize empirical results from:

- Different methods of sample selection and various sample sizes.
- Different topics: disciplines or specialties under study are varied. We must remember that communication patterns and dynamic publishing are very different among disciplines and this can seriously affect the results.
- Different periods in the samples: this is very important given the dynamic nature of the Web, and the changes to which it is subjected (uncontrolled creation, change, and deletion of documents).

### 4.3. Direct query to the database

The third strategy proposed in this working paper consists of asking directly the databases through their search interface, both globally (sectional query) and year by year (longitudinal query).

#### 4.3.1. Sectional query (May 2014)

For a bibliographic database (such as WoS or Scopus), the query is simple and their results easily interpretable; however, in the case of academic search engines (such as MAS or especially GS), this procedure raises a number of unavoidable questions, for example:

- What do the search results obtained in a search engine like Google Scholar refer to? Should all records indexed in the database considered as unique documents?
- Can we trust, to some extent, the results it presents?

*Regarding the first question:*

In the case of Google, the fact that there is no API for Google Scholar, and that Google only displays the first 1,000 results, prevents us from performing large scale empirical studies about the reliability and accuracy of the query commands (especially the "site") command. In the case of Microsoft, Bing Search also retrieves only the first 1,000 results (despite having an API). In the case of MAS (which also offers an API), its functionality is controlled and closer in nature to bibliographic databases, standing midway between a pure academic search engine (Google Scholar) and a pure bibliographic database (WoS).

*Regarding the second question:*

If we were still in 2005-2007, according to Jacsó (2005a; 2008; 2011) we should not trust these results, because at that moment there were paramount mistakes (mainly related to documents with wrong dates of publication and authorship, and duplicates due to not having correctly linked different versions). Today, the answer is probably yes, assuming an error rate that in GS could affect up to 10% of the results. The few empirical studies that have examined these errors have set this error rate below 10%, so our estimate of up to a 10% error rate is likely to be exaggerated. These errors are minimal in traditional bibliographic databases.

To illustrate these differences (and also obtain empirical data to work with), the queries that return the global coverage of both WoS and the academic search engines MAS and GS are offered below. In Figure 6, the query to WoS about the number of registered items from 1700-2014 is shown, obtaining a total of 56,980,000 records.

***Figure 6. Query in WoS about the number of indexed documents (1700-2014)***

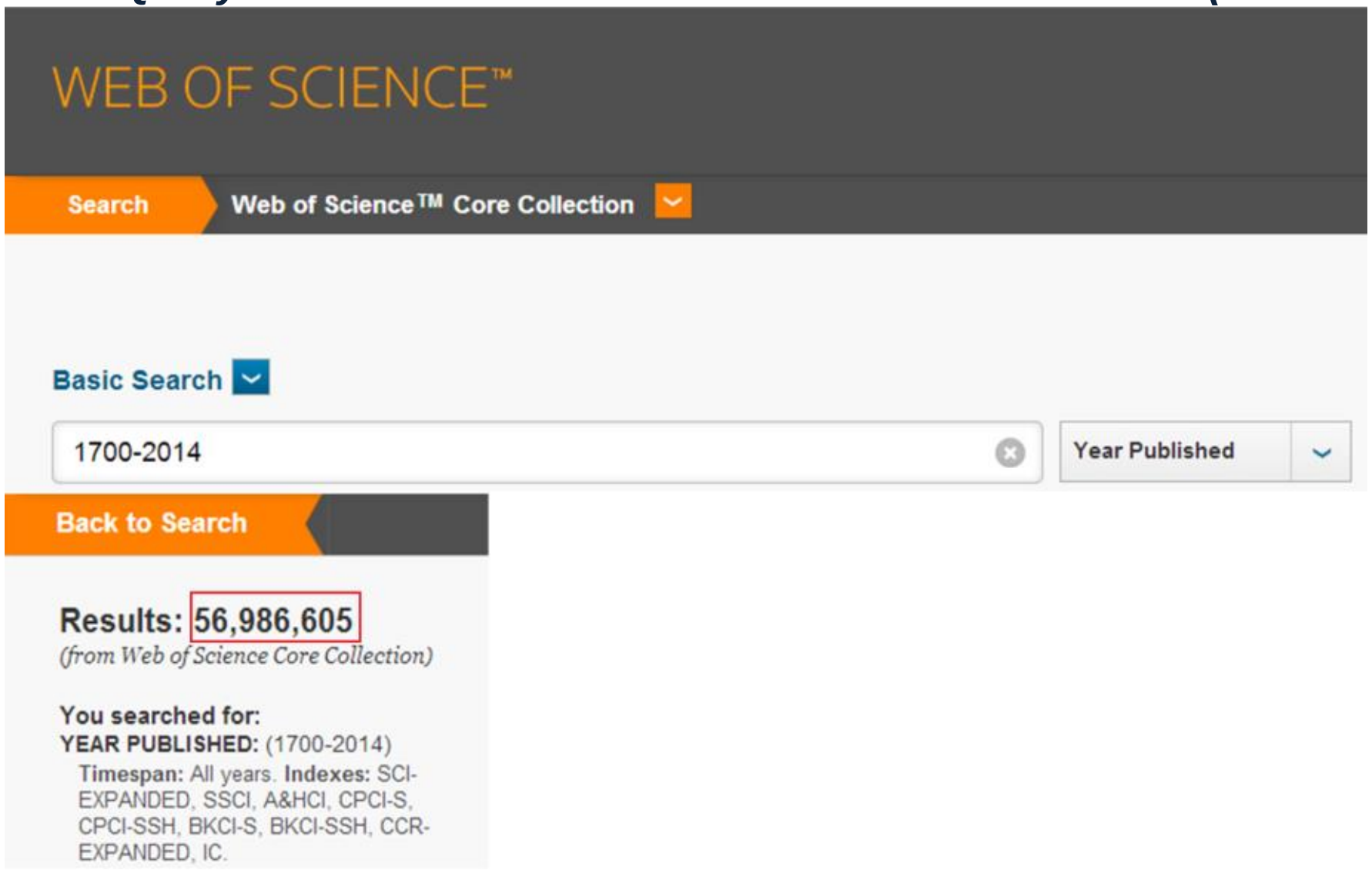

*Source: Web of Science v. 5.13.3*

Nevertheless, if the query is performed for each type of document considered separately, the total differs slightly (Table 3), obtaining 59.5 million records (this is because a document may be classified in more than one type of document).

***Table 3. Number of records according to document type (WoS, 1700-2014)***

| DOCUMENT TYPE | DOCUMENTS | DOCUMENT TYPE | DOCUMENTS |
|---|---|---|---|
| Article | 33,876,866 | Book | 52,395 |
| Meeting abstract | 6,197,232 | Fiction creative prose | 45,357 |
| Proceedings paper | 6,089,411 | Theater review | 31,653 |
| Book review | 3,829,585 | Dance performance review | 21,911 |
| Editorial material | 2,192,341 | Music score review | 18,077 |
| Letter | 2,102,355 | Reprint | 16,093 |
| Note | 1,471,669 | Software review | 15,563 |
| Review | 1,219,612 | Abstract of published item | 13,434 |
| Book chapter | 684,105 | Bibliography | 11,952 |
| News item | 448,039 | Excerpt | 7,396 |
| Poetry | 247,393 | Tv review radio review | 6,881 |
| Correction | 163,682 | Tv review radio review video | 4,791 |
| Correction addition | 157,854 | Script | 2,686 |
| Art exhibit review | 104,853 | Hardware review | 2,540 |
| Biographical Item | 97,023 | Database review | 1,387 |
| Item about an individual | 92,399 | Music score | 1,240 |
| Discussion | 80,425 | Chronology | 1,210 |
| Record REview | 67,044 | Main cite | 13 |
| Music performance review | 61,449 | Meeting summary | 7 |
| Film review | 57,896 | **TOTAL** | **59,495,819** |

*Source: Web of Science v. 5.13.3*

Next, the direct query to MAS is performed (Figure 7) filtering for coverage from 1700 to the present (May 2014).

***Figure 7. Number of records according to discipline in MAS (1700-2014)***

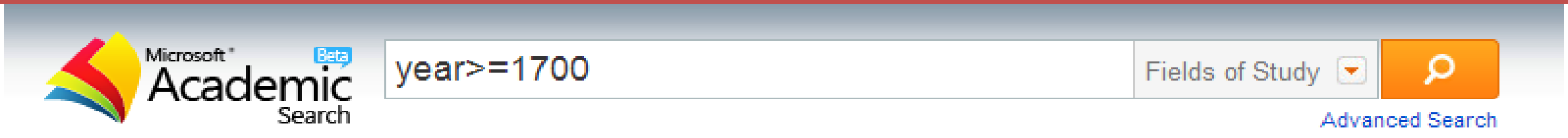

| DISCIPLINE | DOCUMENTS |
|---|---|
| Medicine | 11,576,849 |
| Physics | 4,967,516 |
| Chemistry | 4,380,317 |
| Biology | 3,954,010 |
| Engineering | 3,656,004 |
| Computer Sci. | 3,229,661 |
| Social science | 1,823,846 |
| Art & Humanities | 1,362,358 |
| Geosciences | 1,256,410 |
| Mathematics | 1,143,941 |
| Eco. & Business | 922,478 |
| Material science | 902,532 |
| Agriculture Sci. | 463,560 |
| Environmental Sci. | 449,367 |
| Multidisciplinary | 8,248,035 |
| **TOTAL** | **48,336,884** |

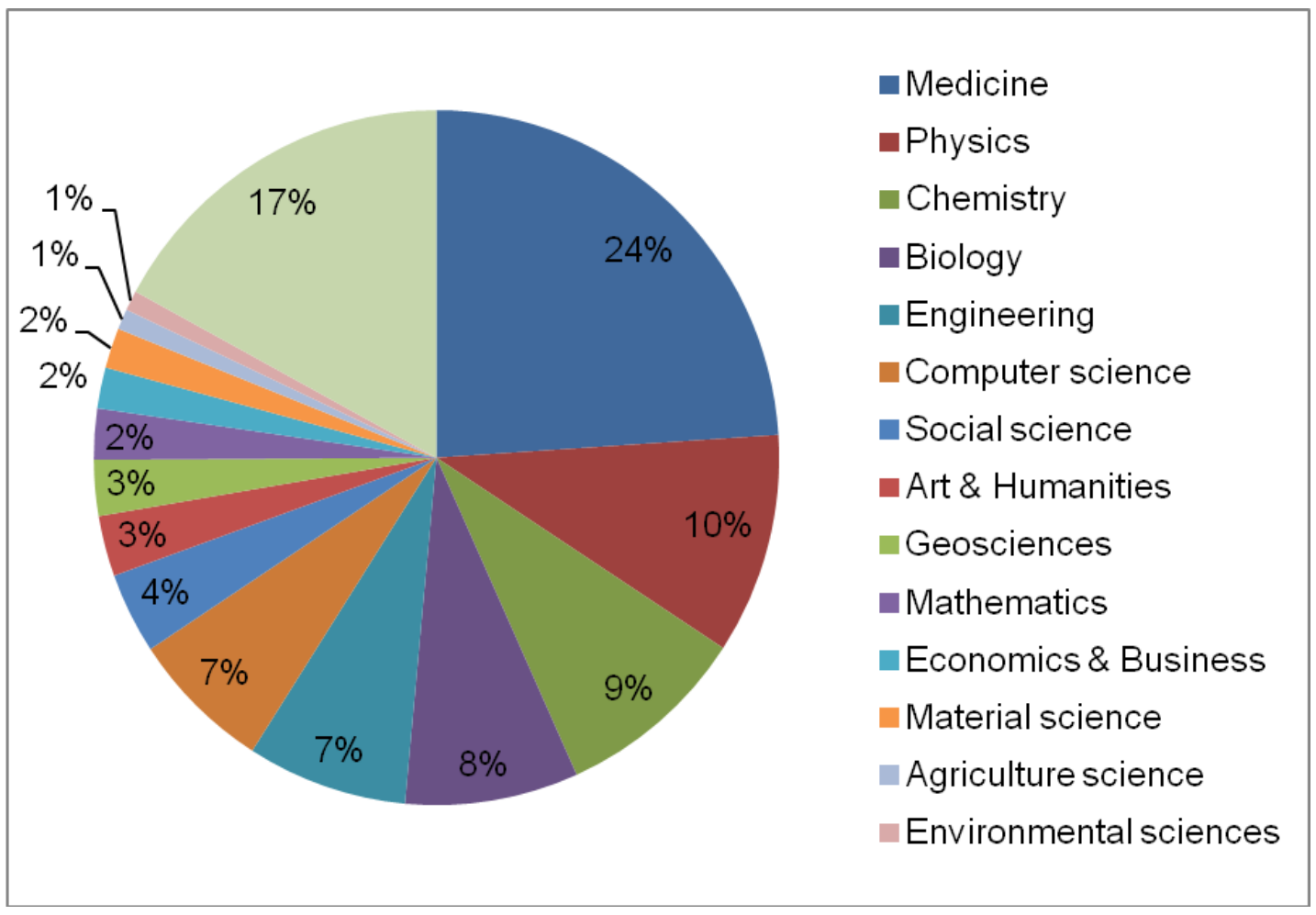

*Source: Microsoft Academic Search*

There is a difference between the direct result (45,970,537 million documents) and that obtained from the summation of the different disciplines (48,336,884) due to the same reason as before: a document can be classified in several different disciplines at the same time.

In the case of Google Scholar, the database allows a temporal query (via custom range option). As with WoS and MAS, we perform a query from 1700 to 2013. Unfortunately, this procedure fails, returning only 596,000 documents. In Table 4, we show some other queries to demonstrate this malfunction.

***Table 4. Malfunction of custom range option in Google Scholar***

| PERIOD | HCE |
|---|---|
| **1700-2013** | 596,000 |
| **1750-2013** | 567,000 |
| **1800-2013** | 552,000 |
| **1850-2013** | 566,000 |
| **1900-2013** | 541,000 |
| **1950-2013** | 617,000 |
| **2000-2013** | 693,000 |

*Source: self-elaborated*

We can observe that the results displayed on Table 3 not only show a low number of results for such a wide timeframe, but also serious inconsistencies. In the time span “2000-2013” the system is retrieving more documents than in longer periods. However, if we execute the query introducing only 1 year in the custom range, the results seem to be more accurate. For example, for the year “1900”, we obtain 141,000 results and for the year “2000”, 2,410,000 results. Therefore, in order to solve this problem, a longitudinal analysis is required.

## 4.3.2. Longitudinal query

The sum of article records from 1700 to 2013 returns 99.8 million records in Google Scholar (59.8 million documents written in English). Comparative data of the three databases (WoS, MAS and GS) are shown in Figure 8.

***Figure 8. Number of documents in GS, MAS and WoS (1700-2014)***

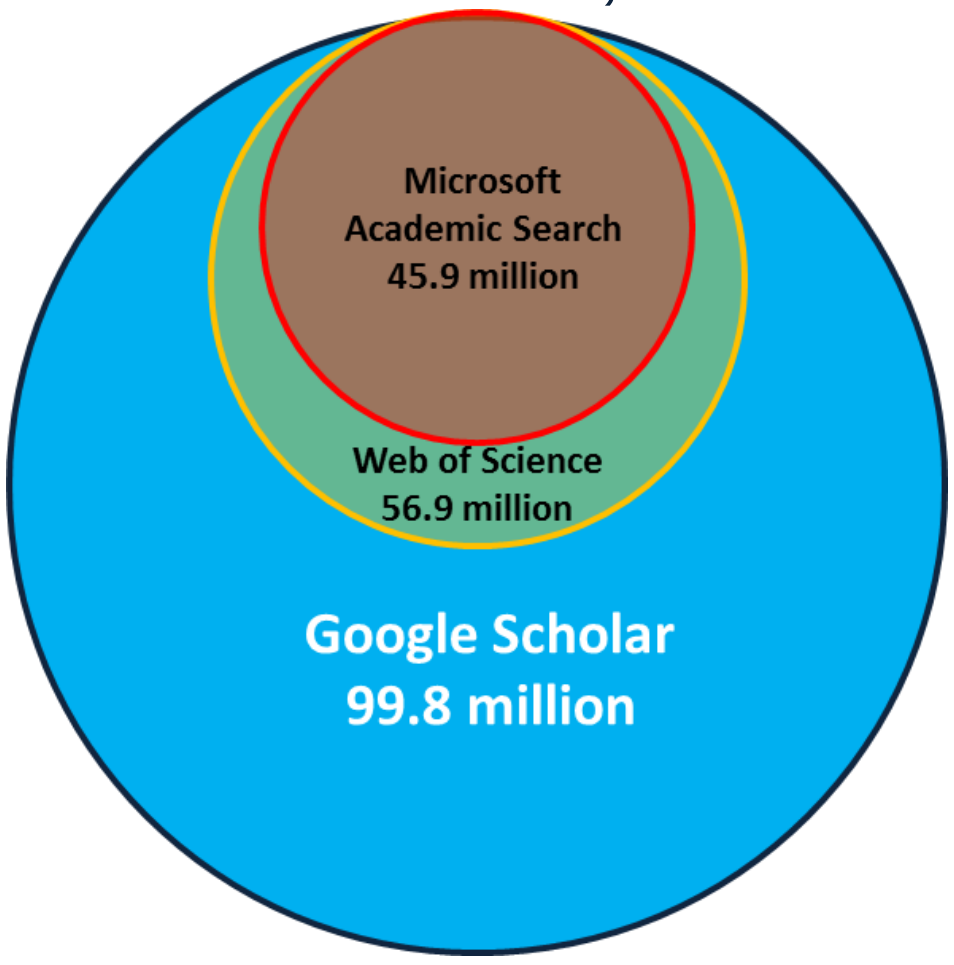

*Source: self-elaborated*
*Circles only represent relative size; intersections do not represent shared coverage*

Next, longitudinal data of the 3 databases (GS, MAS and WoS) from 1800 to 2013 (Figures 8, 9 and 10) is offered[4].

***Figure 8. Google Scholar, Microsoft Academic Search and Web of Science (1800–1899)***

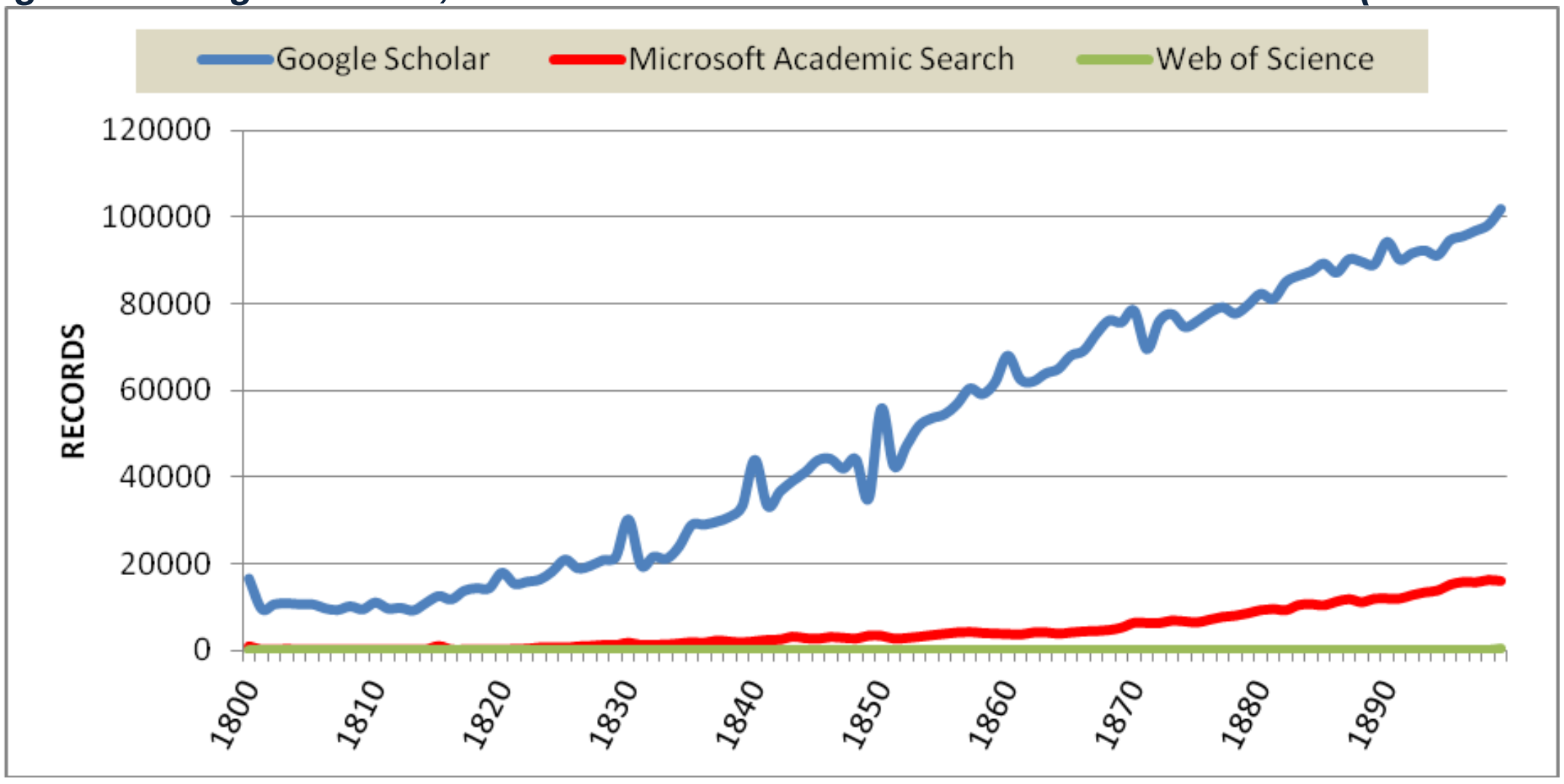

***Figure 9. Google Scholar, Microsoft academic Search and Web of Science (1900–1949)***

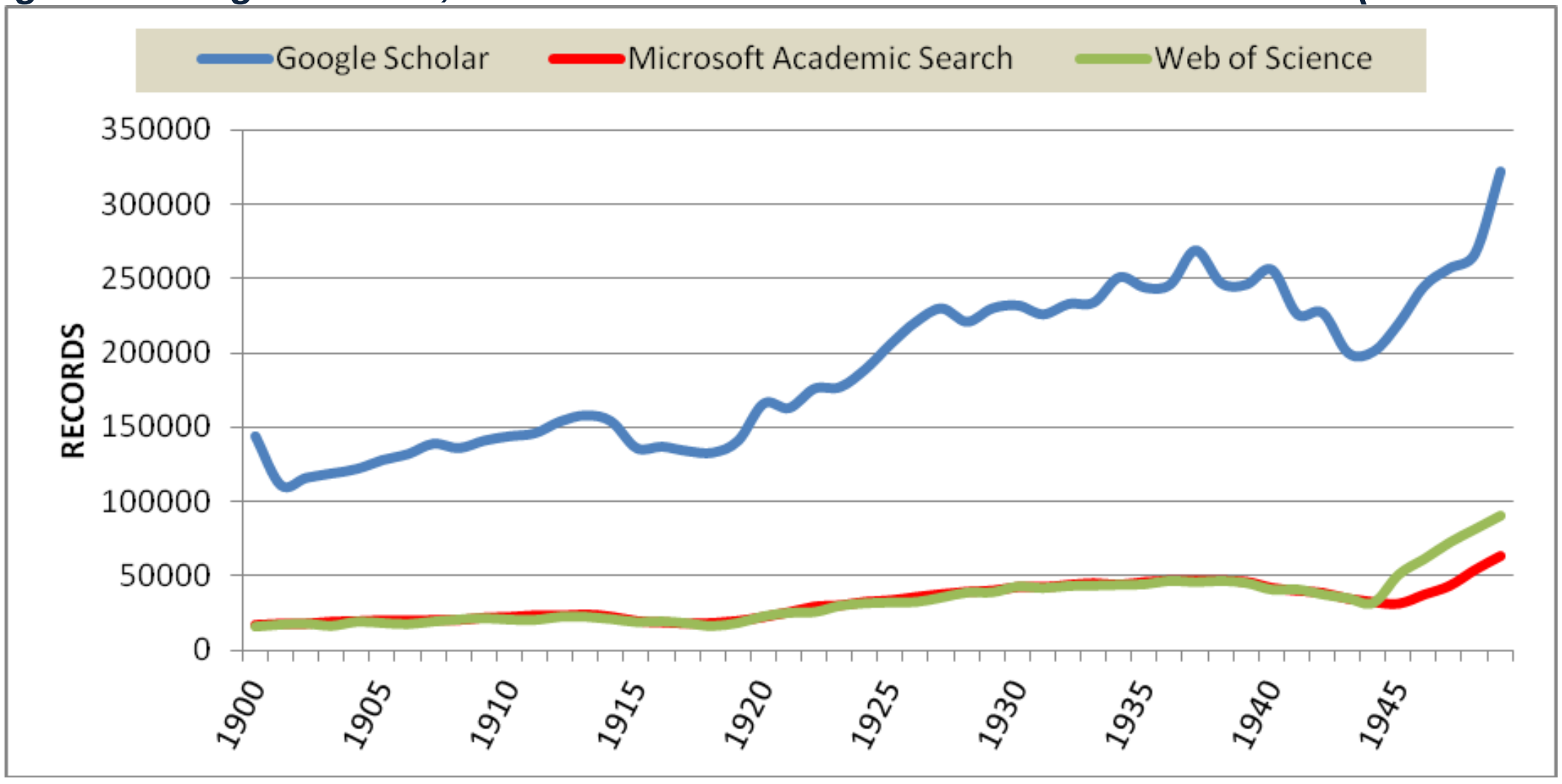

***Figure 10. Google Scholar, Microsoft Academic Search and Web of Science (1950–2013)***

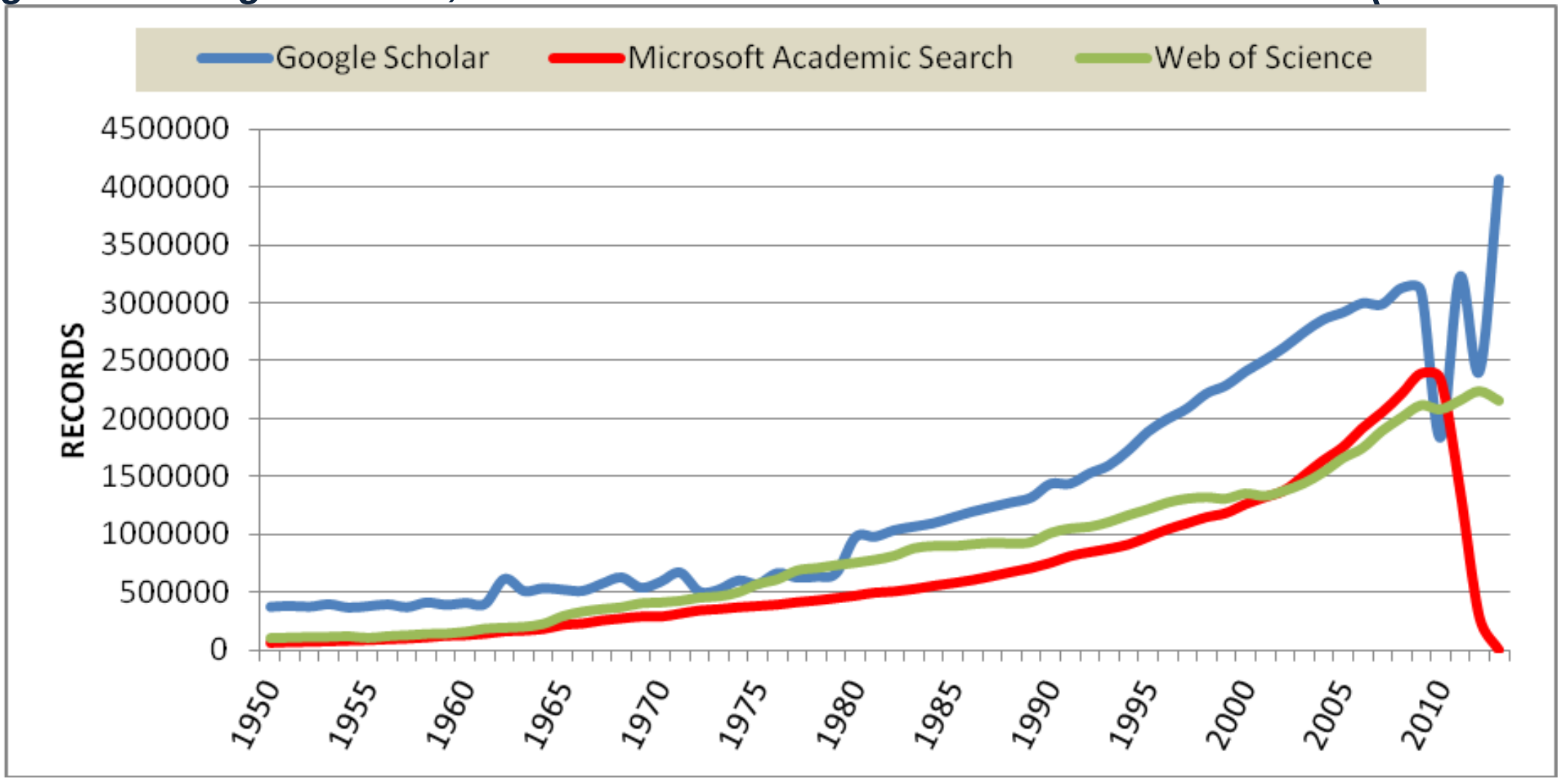

On the one hand, these figures emphasize the primacy of Google Scholar practically during all this period (over 200 years), except in the 1970s, where its performance is similar to that provided by WoS.

The prevalence appears to accelerate again in the last decade of the twentieth century and the first years of the twenty-first, except for the problems identified in 2010 and 2011 (figure 10), probably due to internal changes within the search engine.

Indeed, the problems identified in recent years, very significant on the other hand (a fall of more than 1 million documents from 2009 to 2010, when world output actually accelerates) gives a good account of the dangerous instability of Google Scholar (Aguillo, 2011; Orduña-Malea & Delgado López-Cózar, 2014) and search engine hit count estimates (Jacsó, 2006).

This behavior does not occur (and it would not make sense if it did) on traditional bibliographic databases. The fact that the number of records decreases from year to year obviously does not mean less production in those years, but that the search engine has made internal adjustments, deleting duplicates, fixing bugs, among other technical issues.

On the other hand, these data highlight the similar sizes of WoS and MAS, exemplified in Figure 8, and are consistent with the data offered by Khabsa and Giles (2014). However, between these two products three important differences are emerging:

- MAS collects documents before 1900 unlike WoS (as this database states in its coverage policy).
- MAS drops down since 2010 (Orduña et al, 2014b).
- WoS is growing steadily; in the 1970 it even catches up with Google Scholar.

Finally, Table 5 offers the total count of records grouped by decades since 1950, for each database. Additionally, the relative size of MAS and WoS in relation to Google Scholar (in terms of global size and not on shared records) is offered as well.

For example, in the decade of 2001-2010, the size of WoS was almost two-thirds (62%) the size of Google Scholar, but on the 1970s WoS almost matched the size of Google Scholar (91%). In the case of MAS, 2001-2010 was when it got closest to the size of GS, even surpassing WoS, only to drop in 2010.

***Table 5. Number of records by decade in each database (WoS, 1700-2014)***

| DECADE | GS | MAS | WoS | MAS | WoS |
|---|---|---|---|---|---|
| **1951-1960** | 3,906,000 | 1,006,036 | 1,193,795 | 0.26 | 0.31 |
| **1961-1970** | 5,455,000 | 2,275,739 | 2,919,761 | 0.42 | 0.54 |
| **1971-1980** | 6,467,000 | 3,981,727 | 5,861,577 | 0.62 | 0.91 |
| **1981-1990** | 11,823,000 | 6,107,296 | 8,931,596 | 0.52 | 0.76 |
| **1991-2000** | 19,200,000 | 10,211,009 | 12,119,377 | 0.53 | 0.63 |
| **2001-2010** | 27,730,000 | 18,562,550 | 17,141,610 | 0.67 | 0.62 |
| **2011-2013** | 9,710,000 | 1,692,617 | 6,534,206 | 0.17 | 0.67 |

*Source: self-elaborated*

At this point it should be noted that the sum returning 99.8 million documents in Google Scholar includes both patents and citations. If we exclude these two types of documents from the query, the results fall dramatically to 80.5 million. In Figure 11, we present the results disaggregated by records (80.69%), citations (18.38%) and patents (0.92%), since 1700.

***Figure 11. Composition of Google Scholar results: records, citations and patents***

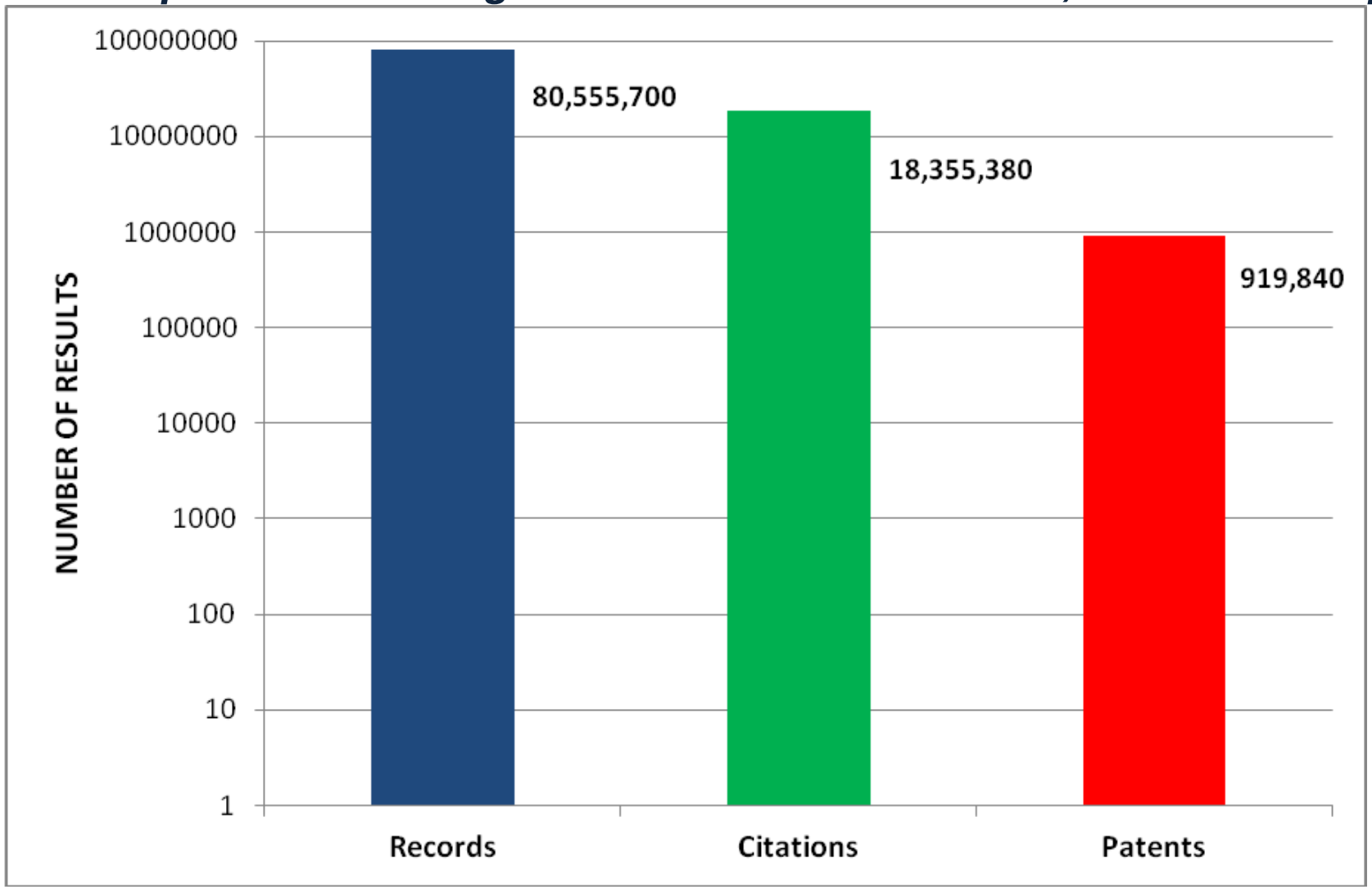

*Source: self-elaborated*

However, these results should be taken with caution, because the hit count estimates in Google Scholar for these queries are not accurate. For the 314 years calculated (from 1700 to 2013), we found inconsistencies up to 122 times between the complete query (records + citations + patents) and the query excluding patents (records + citations), finding more results with the latter than with the former. In short: excluding patents, the system sometimes retrieved more results than in the original complete query.

In Table 6 we show some years where these inconsistencies are present. As is apparent, the problem is accentuated in the last 15 years. Due to the order of magnitude of the results (millions of results since the end of the 20$^{th}$ century), error rates reach unsustainable values, especially if the aim is to make checksums per year, as showed previously in Figure 11.

***Table 6. Inconsistencies in Google Scholar queries for patents and citations***

| YEAR | RECORDS+ CITATIONS+ PATENTS | RECORDS+ CITATIONS | DIFFERENCE |
|---|---|---|---|
| **2013** | 4,070,000 | 4,150,000 | -80,000 |
| **2010** | 1,840,000 | 2,020,000 | -180,000 |
| **2009** | 3,110,000 | 3,230,000 | -120,000 |
| **2007** | 2,990,000 | 3,110,000 | -120,000 |
| **2006** | 3,000,000 | 3,050,000 | -50,000 |
| **2005** | 2,920,000 | 2,950,000 | -30,000 |
| **2004** | 2,860,000 | 2,930,000 | -70,000 |
| **2002** | 2,620,000 | 2,720,000 | -100,000 |
| **2000** | 2,410,000 | 2,550,000 | -140,000 |

*Source: self-elaborated*

As regards citations, the total figure obtained (18,355,380 citations) is higher than we expected. The accuracy is better than that obtained for patents, presenting only 8 errors out of 314 years, and focused in a narrow time span of 20 years: 1969 (difference of -59,000 records), 1970 (-36,000), 1971 (-52,000), 1975 (-19,000), 1976 (-11,000), 1978 (-56,000), 1982 (-10,000) and 1988 (-40,000).

Citations present an additional problem: not all citations are really records Google Scholar hasn't been able to find on the web. In some cases, the same article appears as a record and a citation at the same time. For example, Figure 12 shows a query corresponding to the topic "H Index Scholar". Google Scholar retrieves the same article twice. The first result is considered as a citation, and the second as a regular record. Worst of all, both results count in the global Hit Count Estimate.

***Figure 12. Duplicity in citations on Google Scholar***

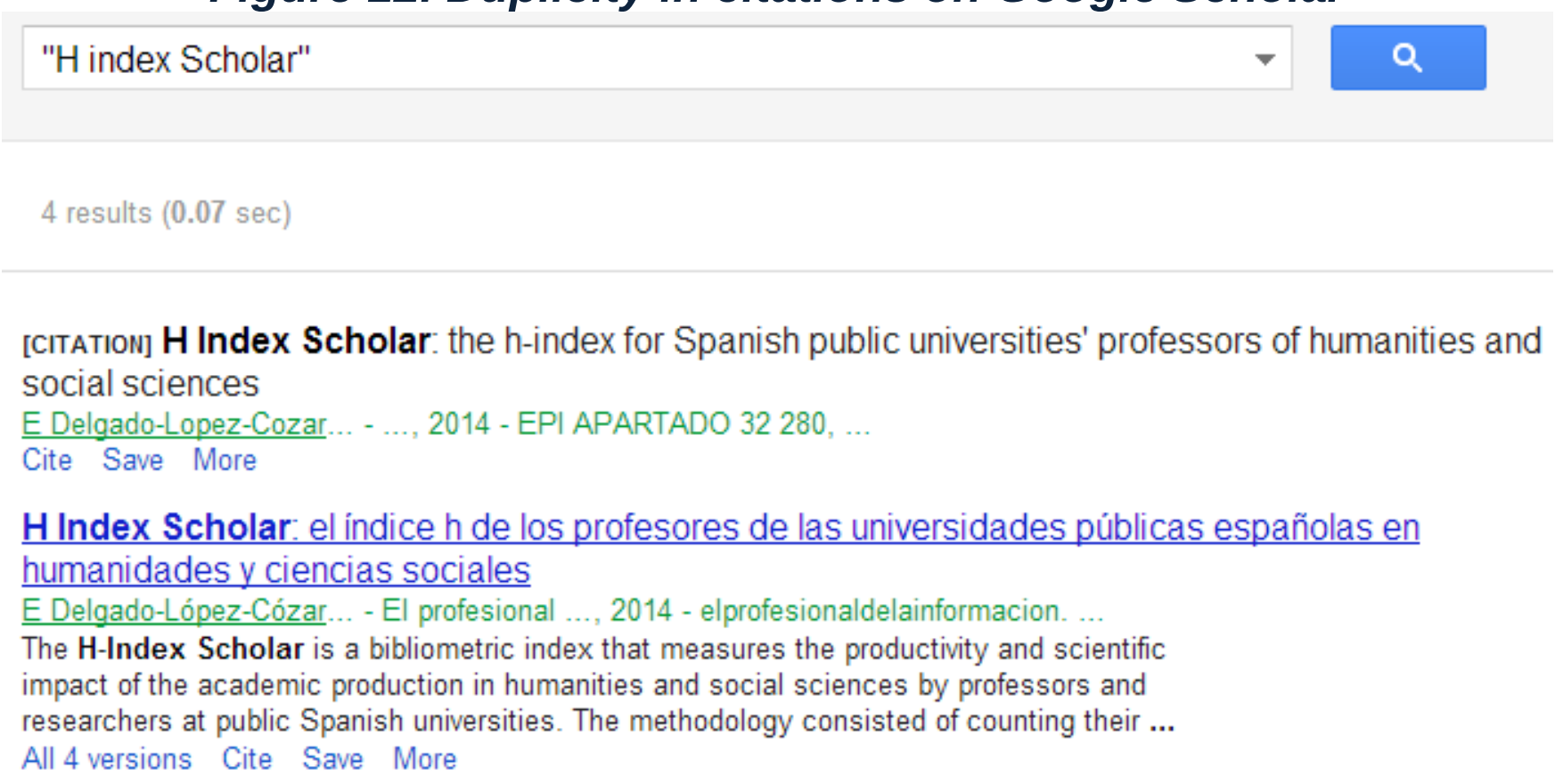

*Source: Google Scholar*

Finally, Google Scholar includes one more type of document apart from the "Articles" category (which is composed of records, citations, and patents): these are the case laws from the Supreme Court of the United States of America, which also include citations (Figure 13).

***Figure 13. Case laws and citations on Google Scholar***

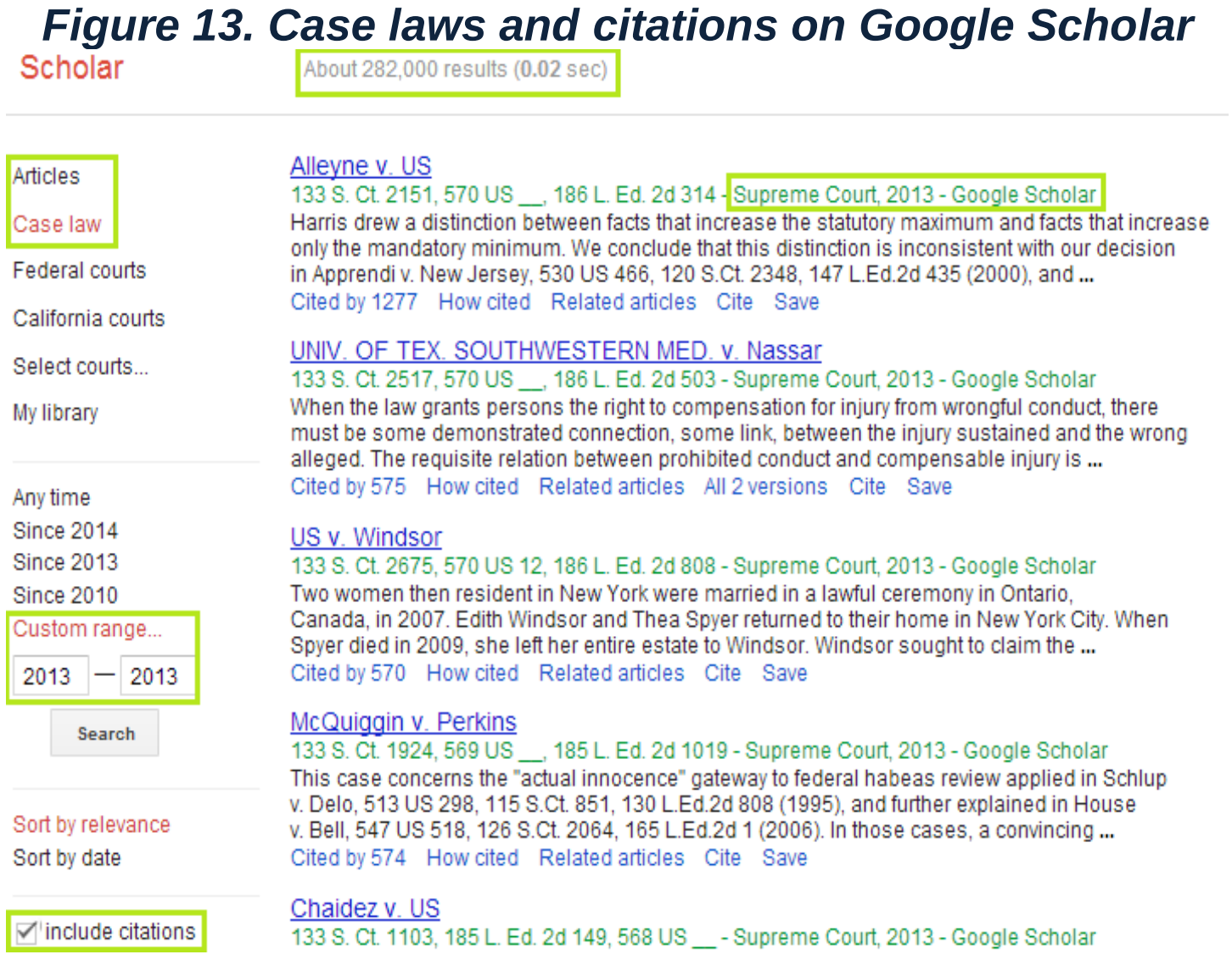

*Source: Google Scholar*

A longitudinal analysis of the number of case laws from 1700 has been performed, in a similar way as for the Articles, both including and excluding citations in the search results. In Figure 14 we can observe the evolution since 1800 (from 1700 to 1799 Google Scholar retrieves only 408 documents).

***Figure 14. Number of Case laws per year (1800-2013)***

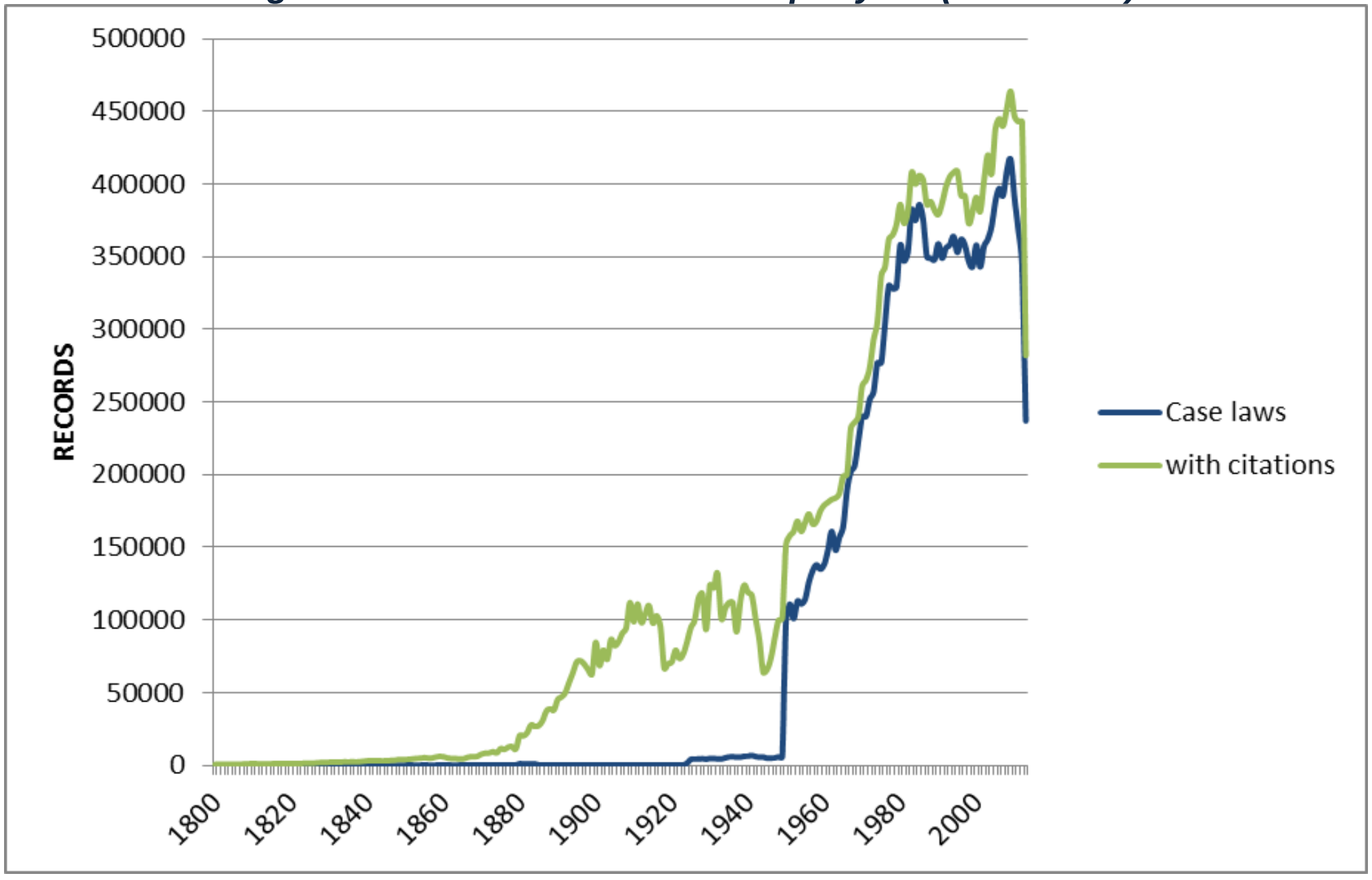

*Source: Google Scholar*

A total of 26,510,689 case laws (31.3%) and citations to case laws (68.7%) have been obtained from 1700. We should highlight the great differences between the number of case laws, and the number of citations to case laws, during the second half of the eighteenth century and the first half of the nineteenth. After that, and until 2013, both types share a very similar behaviour.

If case laws and their citations are included in the calculation of Google Scholar’s total size, the global figure obtained raises to 126,341,609 million documents, a figure about twice the size of WoS (56.9 million).

However, this method also has a number of limitations to consider:

- Errors in rounding results performed by the search engine (we cannot forget that the hit count estimates are, as their name suggests, an estimate).
- The influence of the number of versions (both for records and citations) in the queries.
- Although Articles and Case laws appear as separate categories, are the citations to each one independent? A record marked as citation to Article may be included as a citation to Case law (theoretically, one document can cite both articles and case laws in the reference section).

These shortcomings lead us to consider a priori that this method is probably providing inflated results due to duplicates, versions and upward estimates.

Nevertheless, the figures obtained are unexpectedly lower than those obtained by methods 1 and 2.

## 4.4. Absurd query

Lastly, the fourth method tested consists of applying an absurd query with the purpose of obtaining a hit count estimate about the total size of Google Scholar.

This method is applied under three different approaches: without temporal filter, using the custom range from 1700 to 2013, and finally by means of a longitudinal analysis (year by year).

*a) Without temporal filter*

The query <1 -site:ssstfsffsdfasdfsf.com> is applied to Google Scholar’s Articles category (Figure 15), excluding patents and citations (127,000,000 results), excluding only patents (158,000,000), and including patents and citations (170,000,000).

***Figure 15. Absurd query on Google Scholar***
***Top: excluding patents and citations; middle: excluding patents; down: including all***

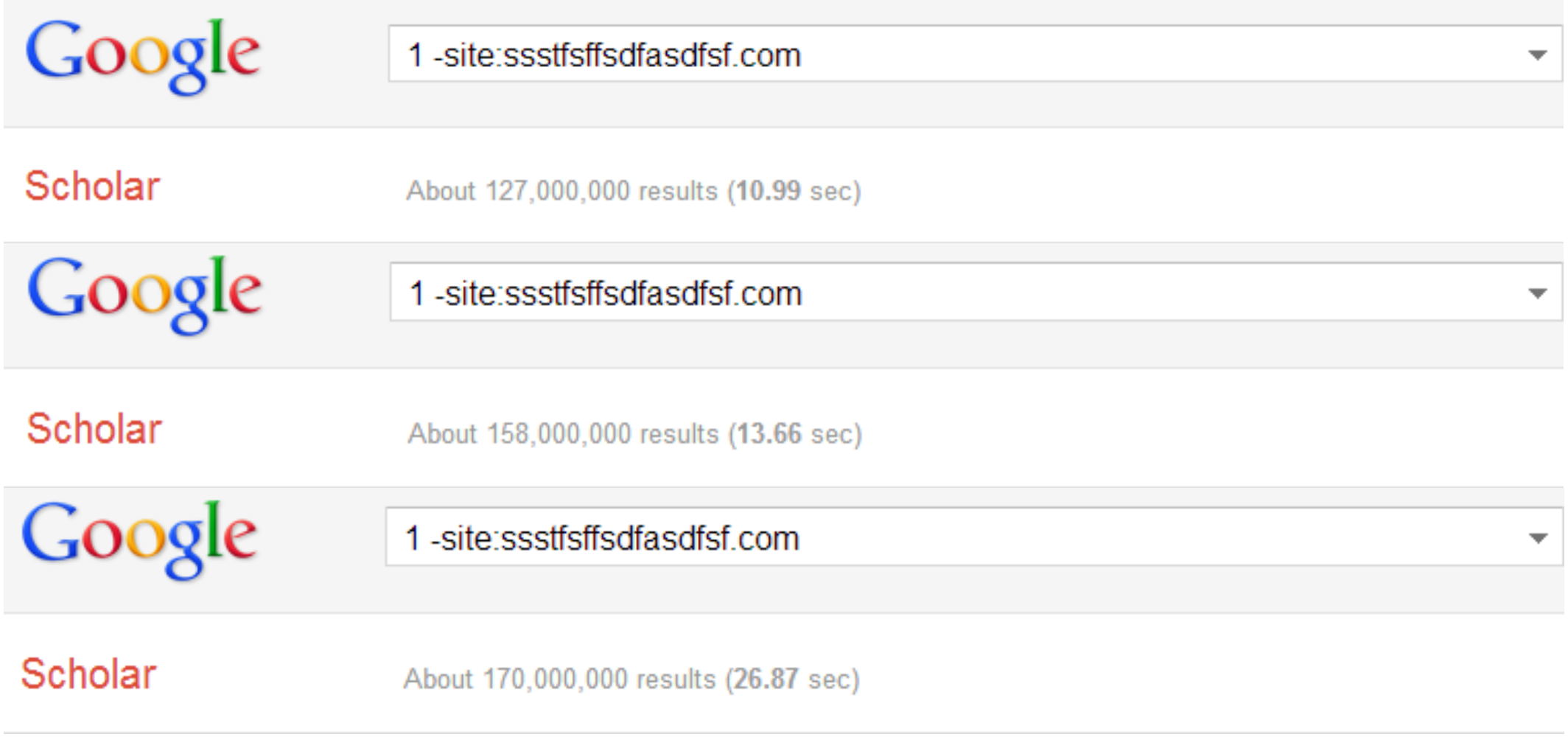

*Source: Google Scholar*

This same query, applied for Case laws (with citations) returns 4,550,000 results, far from the 26.5 million obtained through the longitudinal analysis discussed in the previous method.

The query <a -site:ssstfsffsdfasdfsf.com> has been tested as well, obtaining 102,000,000 (excluding patents and citations) and 154,000,000 (excluding only patents). When trying to include patents and citations (theoretically the query with a higher count), an error message appears informing about technical problems to deliver results (Figure 16).

***Figure 16. Server error with an absurd query on Google Scholar***

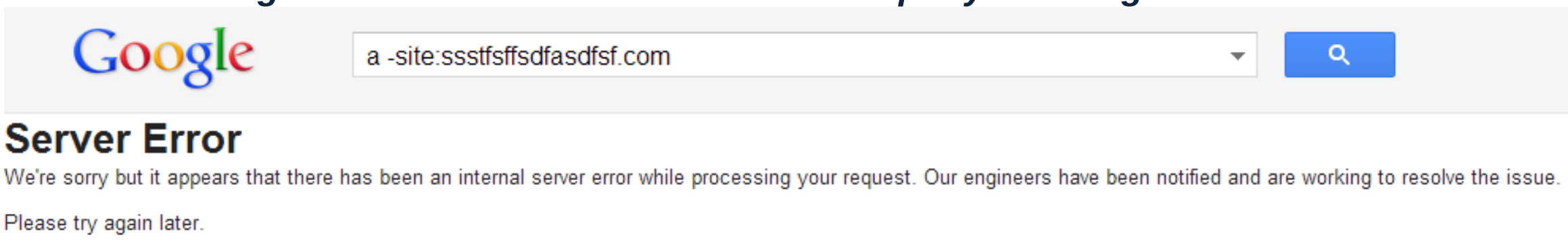

*Source: Google Scholar*

*b) custom range (1700 to 2013)*

In this case, the query <1 -site:ssstfsffsdfasdfsf.com> retrieves 176 million results for Articles and 4.3 million for Case laws. As regards the query <a -site:ssstfsffsdfasdfsf.com>, this time it works returning 160 million articles and 6.8 million case laws..

*c) Longitudinal analysis (1700 to 2013)*

Finally, the absurd query has been performed for each year from 1700 to 2013. The query <1 -site:ssstfsffsdfasdfsf.com> has been selected to do this since it retrieves more results with the procedure "b" than with another query.

The final summation gives an overall of 169.5 million articles and 3.4 million Case laws. These results, although they present different results from those obtained from the longitudinal analysis carried out with the null query used in section 4.3.2, they completely correlate. Pearson correlation (r) for Articles is r = .93 and for Case laws r = .71.

This confirms that Hit Count Estimates from Google Scholar are not useful to get an accurate performance for individual queries, but are useful to make performance comparison and relations.

In Table 7 we summarize the figures obtained from methods 3 and 4, both for Articles and Case laws, with the 3 procedures employed (total query, using the custom range for years, and querying year by year).

***Table 7. Summary of data obtained by the methods of consultation (empty and absurd)***

| ARTICLES | | | |
|---|---|---|---|
| **Absurd query** | | **Empty query** | |
| **Procedure** | **HCE** | **Procedure** | **HCE** |
| Total | 170,000,000 | Total | -- |
| Longitudinal | 169,526,760 | Longitudinal | 99,830,920 |
| Time span | 176,000,000 | Time span | 596,000 |

| CASE LAWS | | | |
|---|---|---|---|
| **Absurd query** | | **Empty query** | |
| **Procedure** | **HCE** | **Procedure** | **HCE** |
| Total | 4,550,000 | Total | -- |
| Longitudinal | 3,422,823 | Longitudinal | 26,510,689 |
| Time span | 4,340,000 | Time span | 629,000 |

*Source: self-elaborated*

We can observe some similarities in the results obtained with the absurd query, both for Articles and Case laws, and independently of the procedure. However, the empty query generates results that are not consistent with those obtained in the previous method (direct query), especially in the case of the longitudinal approach for Articles.

In order to know the reason for the differences between these two methods (both based on querying directly the database), we proceeded to analyse the search results that the absurd query is generating more precisely. Thus we have identified the following weaknesses:

a) The absurd query does not retrieve citations, independently of if this option is checked or not in the search options, both in the case of Articles and Case laws. Conversely, the empty query retrieves citations, as was noted in section 4.3.2. This may explain the differences between these queries in the longitudinal results for Case laws.

b) Hit Count Estimates present serious inconsistencies in the activation / deactivation of the citation inclusion feature. In Figure 17 we display an example of this shortcoming. In the upper figure we display an example of the absurd query, filtered by the year 1840, with the option "Citations" deactivated, which obtains 39 results. At the same time, the bottom figure shows the same query but activating the "Citations" option. As we can see the results obtained in this case are only 9 (although theoretically we should have obtained at least the same as in the other query, or more). Moreover, the system only retrieves 2 documents even though the HCE says there are 9. Although this example deals with Case laws, it is true for Articles as well.

***Figure 17. Hit count estimates inconsistencies in the activation/deactivation of citations***

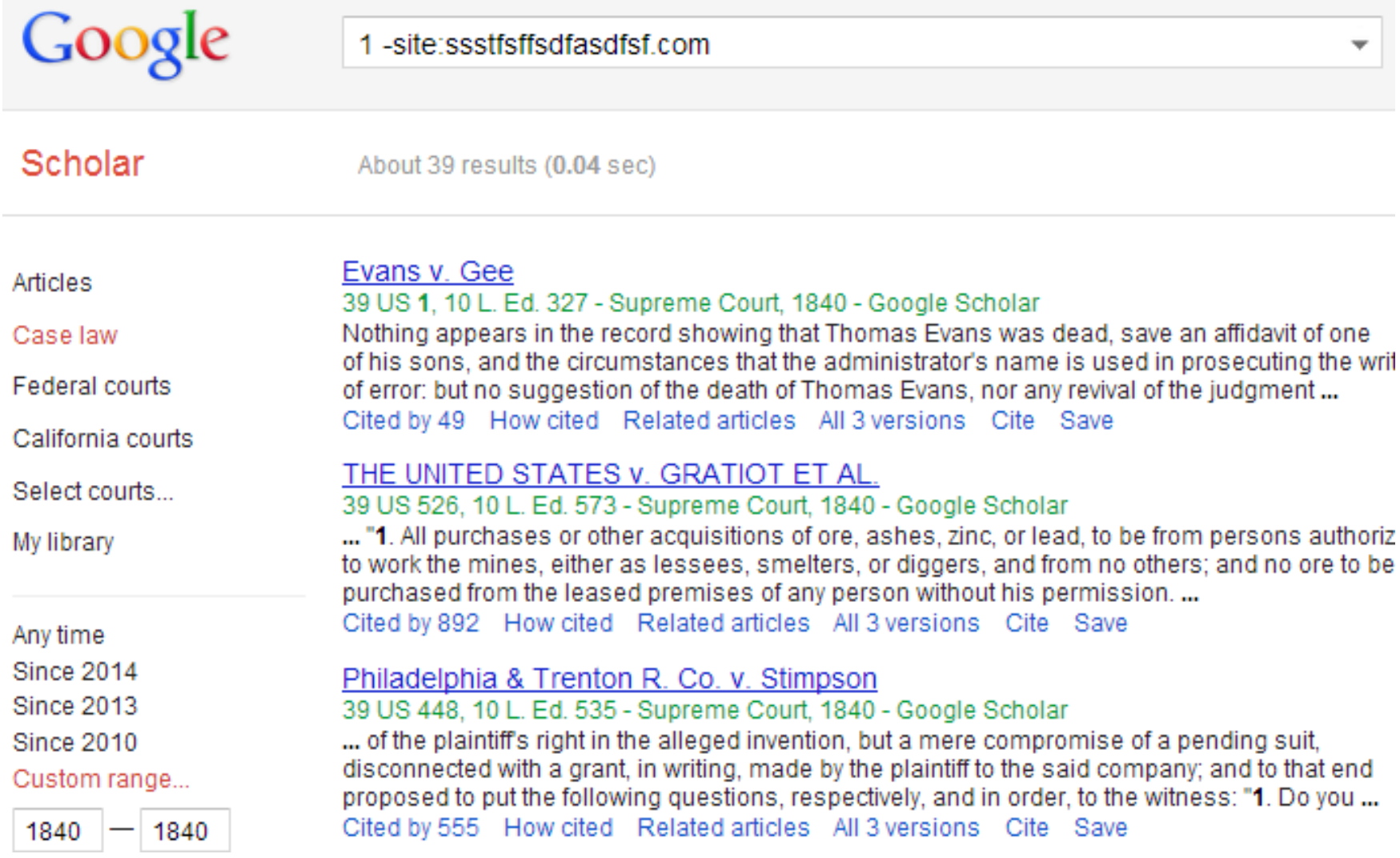

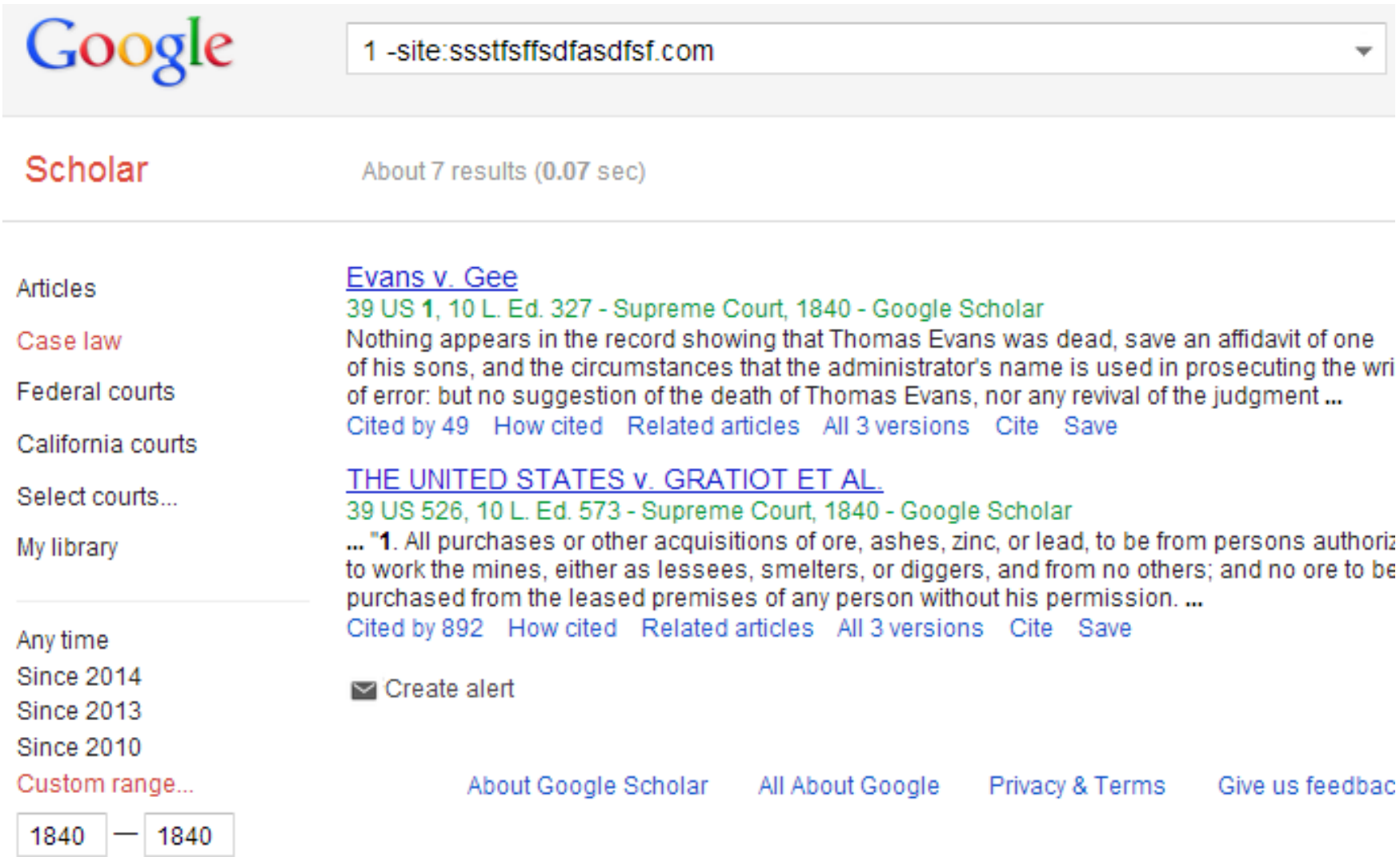

*Source: Google Scholar*

Lastly, we have verified the existence of empty and false SERPs (Search Engine Results Pages). In Figure 18 we display an example of an application of the absurd query, for any given year. In the upper figure we can observe that the system retrieves a Hit Count Estimate of 132, but the 6th SERP is empty. If we set the system to retrieve up to 20 results per SERP (the maximum allowed in GS), the 6th SERP should show results 101th to 120th, and never an empty result. Moreover, if we click on the 15th SERP (bottom figure), the system not only still retrieves an empty SERP but the HCE increases as well (521). It should be noted that the longitudinal analysis has been performed using the first SERP.

***Figure 18. Empty and false SERPs in Google Scholar***

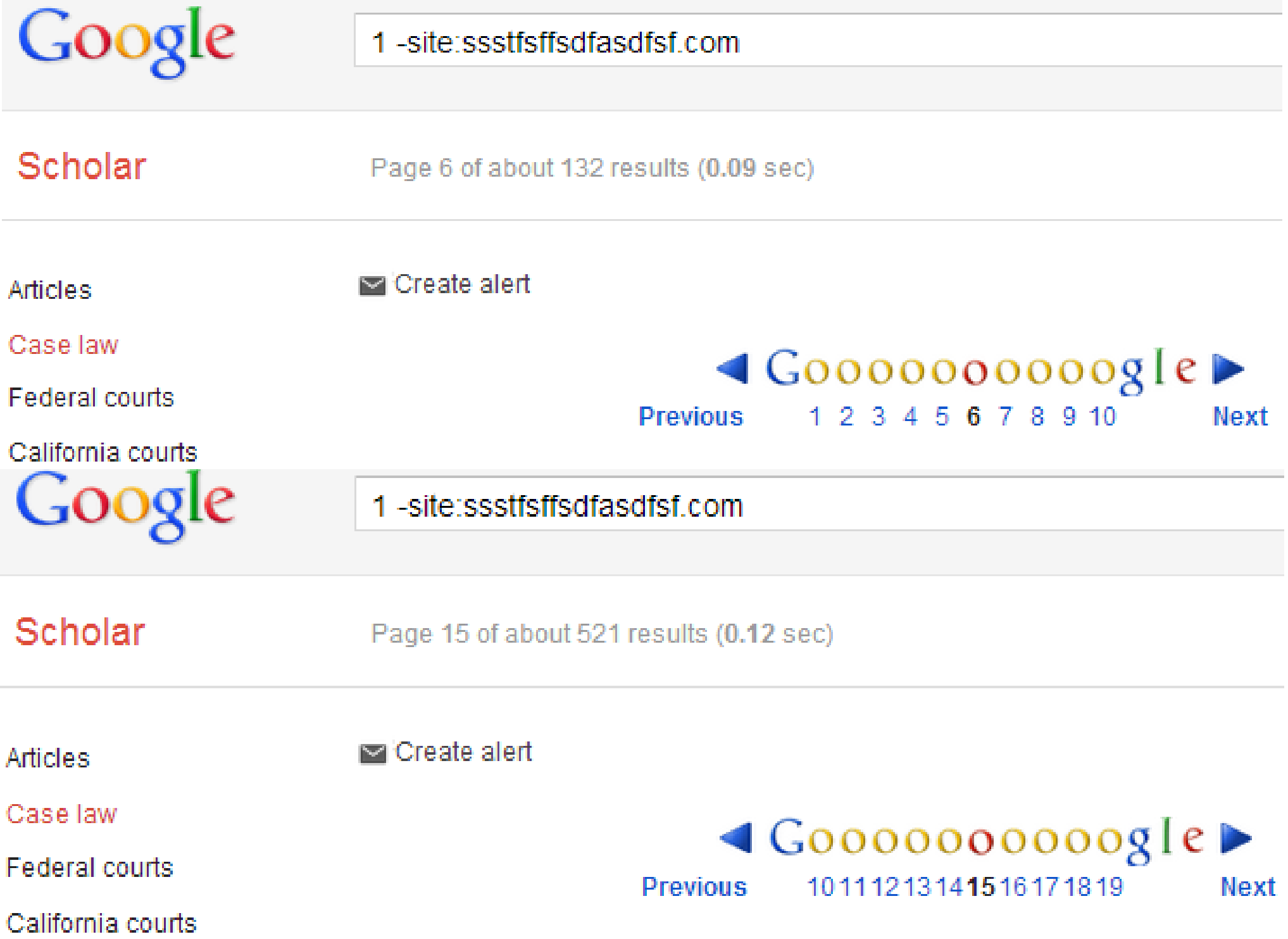

*Source: Google Scholar*

Since this method of absurd query is more accurate than it seems at first because the search engine is forced to check the entire database to answer the query, as the time responses are suggesting (See Figure 15), these shortcomings are, as of yet, unexplained.

It is possible that the system goes into a loop when trying to answer a query of this type, but it is more surprising that the final figures provided seem logical and coherent, and close to those achieved by other methods, unlike what happens with the empty query method.

## 5. CONCLUSIONS

The main objective of this study was to measure the size of Google Scholar via four different techniques, discussing the advantages and disadvantages of each method and the disparity of results. Table 8 summarizes the results obtained for each method.

***Table 8. Summary of Google Scholar size estimates***

| METHOD | GS SIZE ESTIMATE | COMMENTS |
|---|---|---|
| **A. Khabsa & Giles** | 152.7 million | The authors estimate 99.3 million for English contents without patents. If we consider that English constitutes approximately 65% of contents we obtain this figure. |
| **B. Empirical data** | 171 million | This method assumes GS triples the size of WoS |
| **C.1. Empty query (custom range)** | 1.2 million | This method applies an empty query in setting the custom range from 1700 to 2013. |
| **C.2. Empty query (longitudinal)** | 126.3 million | This method applies an empty query in a longitudinal analysis from 1700 to 2013 including Articles (99.8 millions) and Case laws (26.5 millions). |
| **D.1. Absurd query (total)** | 174.5 million | 170 million Articles and 4.5 million Case laws |
| **D.2. Absurd query (custom range)** | 176.8 million | This method applies an absurd query setting the custom range from 1700 to 2014, both for Articles (170 millions) and Case laws (6.79 millions) |
| **D.3. Absurd query (longitudinal)** | 172.9 million | This method applies an absurd query in a longitudinal analysis from 1700 to 2013, obtaining 169.5 million Articles and 3.4 million Case laws. |

*Source: self-elaborated*

Method A (taking apart the identified problems about the use of the Lincolm-Petersen method, the academic size outside Google Scholar or the biased sample) poses a priori some methodological problems, because it parts from a false axiom: inferences should not be made from the comparison of databases with very different characteristics.

The database usually employed to make comparisons is the Web of Science, but it has, among others, two major flaws: the documents it indexes are mostly written in English, and most of them are journal articles.

The well-known problems of WoS can also be applied to MAS, from which the authors extract the sample. The main concerns in this sense are the following:

a) The “Citing documents” can be in all languages. The authors identify that 98% of citing papers in MAS are written in English (a correction is applied instead of eliminating non-English documents), but this percentage in Google Scholar must be lower, and it is not indicated in the estimation of the size of Google Scholar. Probably all documents written in languages other than English should have been avoided if the estimation of just the “English academic web” was the objective.

b) The sample is composed by articles, whereas citing documents are diverse in their typology. If the target of this research had been calculating the number of “articles” included on Google Scholar, then this procedure would have been appropriate (after the elimination of citing documents other than articles). If the purpose was to calculate the size of the entire database, a sample composed uniquely by articles is not representative of Google Scholar.

For this reason, we believe that on the one hand, the calculation is oversized with respect to the language (the 99.3 millions obtained cover more than English Article contents), and undersized with respect to the document types. Moreover, patents and case laws (integrated in the Google Scholar database) should be added in the final count.

Nevertheless, the research design performed by the authors is novel and brilliant. It is based on the gathering of citing documents to a sample of cited articles. This procedure has several advantages (such as that the search engine is forced to query its entire database to find all documents that match with a citation to any of the documents of the sample). Nonetheless, there are types of documents on Google Scholar (such as syllabi, conferences, teaching material, etc.) that maybe do not cite any other document (and they will never be cited as well), that are not considered in this method.

Method B (making estimates based on the comparative differences between databases), provides an approximate size for Google Scholar of 171 million records. This method has the advantage of not being affected by comparisons with other biased databases, although the estimate is very rough and imprecise as it is synthesized from diverse empirical results. Despite this, the results obtained are close to those obtained in the method A.

A priori, method C seemed to be more accurate, being based on direct querying of Google Scholar, but the results are unexpectedly more distant from those of other methods (Case laws are overrepresented and Articles are underrepresented), especially method C.1, which is discarded completely.

In the case of method C.2 (longitudinal query), the problems stem from the following issues:

- Lack of precision: the extent to which the search engine returns data for our query that corresponds with the reality of its cataloged universe. We rely on hit count estimates (Google explicitly states “about xxx results”), which are affected by unknown rounding routines.

- Lack of reliability: the extent to which the search engine returns, from our query, what we really want to measure, i.e., the number of unique records indexed in Google Scholar.

In that sense, if we consider up to a 10% of errors (related with cataloging issues, duplicate records, etc.), and errors related with the accuracy and reliability, the actual size of Google Scholar may be even lower according to this method, around 114 million records.

These results are completely unexpected, especially because when using this method we are including patents and the Case law category (with their corresponding citations). Both types of documents are excluded from both the methods A and B. Therefore we expected higher results, not the opposite.

It is probable, however, that:

- Method A gives inflated results if we assume that the 99.3 million gathered are including not only English documents, and not only article-type documents.
- Method B gives inflated results for Google Scholar because these comparisons focus on periods where the supremacy of Google Scholar was higher. This is because GS may triple WoS if we consider the whole timeframe, but there are periods in which this is not true (Figures 8-10).

Finally, method D, though simple, produces similar results to those obtained by methods A and B. This procedure, though it is closer to method C, presents results which are significantly different to those of method C. The Hit Count Estimates, which skew both methods, may influence each method differently, or it is also possible that the absurd query does not work properly for some reason.

In this sense, we have checked that the absurd query does not retrieve citations (independently of if this option is checked or not in the search options), and it also creates empty and false SERPs. Moreover there is a dysfunction between the inclusion of Citations and Patents and the HCE obtained (in this case both for the Articles and Case law categories, and both for empty queries and absurd queries).

The underlying questions here are the following: what should we have to doubt more?, the reliability of Google Scholar's HCEs (assuming 10% error in the results)?, the estimate from MAS using the Lincolm-Petersen method?, or the estimate based on a correction factor?

Most surprising of all is that even though all methods seem invalid for various and diverse reasons, all return similar results (except Method C). Probably a sensible estimate, observing all the results obtained and taking into account a 10% of possible internal errors, would be a total of around 160 million unique records (without considering different versions for the same record).

The information policies of GS ("no comments" is the house brand) encourage speculation and force researchers to make conjectures about the real size of this database (and its entrails). A figure that otherwise surely would be easy to determine by their technical and manager staff simply by pressing a key on their computer at the office.

Logically this matter should be solved by simply asking this information to Google, and to the people directly responsible for Google Scholar. Their answer would avoid all our concerns, efforts and resources dedicated to finding this sort of “golden fleece” that this issue has become.

Although it seems impossible that Google will publish this information (at least on a short term), we wonder if anyone can "press the button" and tell us what the size of Google Scholar is. Perhaps even Google Scholar does not know this “number”… a number that approximately represents the online scientific heritage circulating at present.

## Notes

1. The option Custom range appears after a query is submitted, in the search box of Google Scholar (not before). Moreover, we can execute this query directly on the browser via http as well. Once we obtain the first results via hit count estimates, we can generate new queries without introducing any keyword in the search box, and only selecting the time span required.
2. https://www.worldcat.org
3. http://futur.upc.edu
4. Web of Science does not provide data until 1898.

## APPENDIX I. CATALOGUE OF EMPIRICAL WORK RELATED TO THE SIZE OF GOOGLE SCHOLAR

| AUTHORS | SAMPLE | UNIT | GS/GSM/GSC | WoS | MAS | SCO | PUB | PSY | ERIC | SSCI | SJR | JCR | JSC | ECO | CAB | SCI | CA | GB | KoMCI | KMS | CINAHL |
|---|---|---|---|---|---|---|---|---|---|---|---|---|---|---|---|---|---|---|---|---|---|
| **KHABSA & GILES (2014)** | 150 English written documents from MAS; 10 of the most cited documents in each of the 15 fields are randomly sampled. | Citations | 86,870 | | 41,778 | | | | | | | | | | | | | | | | |
| **KHABSA & GILES (2014)** | 1,500 documents from MAS; 100 documents belonging to each field, with at least 1 citation.(n=114 million) | Documents (million) | 99.3 | 49 | 48 | | | | | | | | | | | | | | | | |
| **ABDULLAH & THELWALL (2014)** | Books (n= 1,357) and citations (n=2,254) of Malaysian books in AHSS disciplines. | Documents | 499 | | | | | | | | | | | | | | | 314 | | | |
| | | Documents (%) | 37 | | | | | | | | | | | | | | | 23 | | | |
| | | Citations per book | 1.67 | | | | | | | | | | | | | 1.61 | | | | | |
| **WINTER & ZADPOOR & DODOU (2014)** | Number of citations on to Garfield for WoS and GS | Citations | 1,231 | 607 | | | | | | | | | | | | | | | | | |
| | | Unique citations | 703 | 153 | | | | | | | | | | | | | | | | | |
| | | (%) | 90 | 48.2 | | | | | | | | | | | | | | | | | |
| **HALEY (2014)** | 50 Top economics finance journals were selected and then scored using both GS and MAS using the PoP software (1993–2012) | Average h-index | 154.40 | 78.98 | | | | | | | | | | | | | | | | | |
| | | Average h-index | 267.36 | 125.72 | | | | | | | | | | | | | | | | | |
| | | Average AWCR | 9834.00 | 2740.87 | | | | | | | | | | | | | | | | | |
| | | Average e-index | 186.21 | 81.00 | | | | | | | | | | | | | | | | | |
| **ORDUÑA-MALEA & DELGADO LÓPEZ-CÓZAR (2014)** | World weekly (average) size and monthly growth rate per source | Weekly size (average) | 68,545,750 | 27,904,896 | 28,113,479 | | | | | | | | | | | | | | | | |
| | | Monthly growth rate (%) | 11.15 | 0.37 | 0.41 | | | | | | | | | | | | | | | | |
| **ORTEGA & AGUILLO (2014)** | Analysis of the 771 personal profiles appearing in both the MAS and the GSC. | Documents (%) | 158.3 | | 89.5 | | | | | | | | | | | | | | | | |
| | | Citations (%) | 327.4 | | 76.7 | | | | | | | | | | | | | | | | |
| | | % | 155.8 | | 72.1 | | | | | | | | | | | | | | | | |
| **CARDENAS & UDO (2013)** | Knowledge management (KM) articles published between 1993 and 2012 | KM papers | 33,600 | 9,887 | | | | | | | | | | | | | | | | | |
| | | KM papers (%) | 77.26 | 28.59 | | | | | | | | | | | | | | | | | |
| | | KM papers in USA | 12,434 | 2,084 | | | | | | | | | | | | | | | | | |
| | | KM papers in USA (%) | 80.65 | 14.35 | | | | | | | | | | | | | | | | | |
| **ADRIAANSE & RENSLEIGH (2013)** | African scholarly environmental sciences journals the period 2004-2008 (n=3,199) | Citations | 2,715 | 2,740 | | 2,192 | | | | | | | | | | | | | | | |
| | | Overall coverage (%) | 84.9 | 85.7 | | 68.5 | | | | | | | | | | | | | | | |
| | | Inconsistencies | 448 | 165 | | 14 | | | | | | | | | | | | | | | |
| | | (%) | 14 | 5.2 | | 0.4 | | | | | | | | | | | | | | | |
| **CABEZAS-CLAVIJO & DELGADO LÓPEZ-CÓZAR (2013)** | Most relevant journals and researchers in the field of intensive care medicine. | Average Journal H-index | 36 | 28 | | 32 | | | | | | | | | | | | | | | |
| | | Average Author H-index | 29 | 23 | | 25 | | | | | | | | | | | | | | | |
| **DELGADO LÓPEZ-CÓZAR & CABEZAS-CLAVIJO (2013)** | Nº Journals indexed GSM, JCR and SJR | Journals | 40,000 | | | | | | | | 19,708 | 10,677 | | | | | | | | | |
| **DELGADO LÓPEZ-CÓZAR & REPISO (2013)** | Sample of journals from the field of communication studies indexed in three databases. | Number of communication journals covered | 277 | 106 | | 167 | | | | | | | | | | | | | | | |
| **HUH (2013)** | Citation indicators of the Korean journal of urology before and after 2010 | Citations | 428 | 86 | | 134 | | | | | | | | | | | | | | | |
| | | (%) | 44.70 | 9 | | 14.02 | | | | | | | | | | | | | 207 | 101 | |
| **ZARIFMAHMOUDI & KIANIFAR & SADEGHI (2013)** | Citations to 2011-2012 articles of Iranian Journal of Basic Medical Sciences (IJBMS) in three resources | Citations | 59 | 20 | | 30 | | | | | | | | | | | | | | | |
| | | (%) | 44.56 | 29.59 | | 25.85 | | | | | | | | | | | | | | | |
| | | Unique citations | 40 | 2 | | 6 | | | | | | | | | | | | | | | |
| | | GS and WoS = 11; GS and SCO = 17; SCO and WoS = 16; GS and SCO and WoS =9 | | | | | | | | | | | | | | | | | | | |
| **AMARA & LANDRY (2012)** | Data of scholars of Canadian business schools, used jointly with data extracted from the WoS and GS databases . | Contributions record | 21.56 | 5.08 | | | | | | | | | | | | | | | | | |
| | | Citations record | 271.53 | 50 .85 | | | | | | | | | | | | | | | | | |
| | | Hirsch h-index | 4.57 | 1.87 | | | | | | | | | | | | | | | | | |
| | | Proportion of contributions | 72.21 | 27.79 | | | | | | | | | | | | | | | | | |
| | | Proportions of citations | 69.28 | 30.72 | | | | | | | | | | | | | | | | | |
| **DE GROOTE & RASZEWSKI (2012)** | 100 articles from the publications of 30 College of Nursing faculty (n=3,000) | Articles | 927 | 795 | | 974 | | | | | | | | | | | | | | | |
| | | % | 34.38 | 29.49 | | 36.13 | | | | | | | | | | | | | | | |
| | | H-index | 316 | 215 | | 271 | | | | | | | | | | | | | | | |
| | | Aggregated h-index | WoS, Scopus, GS, and CINAH =131; WoS, Scopus and CINAHL=339; WoS and Scopus = 326 | | | | | | | | | | | | | | | | | | |
| | | Citations | 3,492 | 1,406 | | 2,437 | | | | | | | | | | | | | | | 966 |
| | | % | 42.10 | 16.93 | | 29.34 | | | | | | | | | | | | | | | 11.63 |
| | | Unique citations | 1,312 | 93 | | 250 | | | | | | | | | | | | | | | 273 |
| | | % | 68.05 | 4.82 | | 12.97 | | | | | | | | | | | | | | | 14.16 |
| **GIL ROALES-NIETO & O'NEILL (2012)** | Articles published in IJP&PT and cited (2001-2010) | Articles | 238 | 238 | | 231 | | | | | | | | | | | | | | | |
| | | Cited Articles | 208 | 171 | | 167 | | | | | | | | | | | | | | | |
| | | (%) | 87 | 72 | | 72 | | | | | | | | | | | | | | | |
| **LASDA BERGMAN (2012)** | Top five journals ranked highest in overall quality by the 556 faculty members surveyed (2005) | Citations | 3,272 | 1,741 | | 2,126 | | | | | | | | | | | | | | | |
| | | Unique citations | 1,904 | 197 | | 339 | | | | | | | | | | | | | | | |
| | | Overload (%) | 44.2 | 4.6 | | 7.79 | | | | | | | | | | | | | | | |
| | | GS and WoS = 81 (1.9%); GS and SCO = 324 (7.5%); SCO and WoS = 502 (11.7%); GS, SCO and WoS = 961 (22.3%) | | | | | | | | | | | | | | | | | | | |
| **MIRI & RAOOFI & HEIDARI (2012)** | 104 articles of Hepat Mon published in 2009 and 2008 which had been cited in 2010 in three databases including WoS, SCO and GS. | Articles | 100 | 87 | | 91 | | | | | | | | | | | | | | | |
| | | (%) | 91 | 83.65 | | 87.5 | | | | | | | | | | | | | | | |
| | | Citations | 85 | 69 | | 86 | | | | | | | | | | | | | | | |
| **ZARIFMAHMOUDI & SADEGHI (2012)** | Citations to 100 articles of Iranian Journal of Nuclear Medicine (IJNM) from 2006-2012 in SCO and GS. | Articles | 100 | | | 99 | | | | | | | | | | | | | | | |
| | | Coverage (%) | 100 | | | 99 | | | | | | | | | | | | | | | |
| | | Unique citations | 18 | | | 9 | | | | | | | | | | | | | | | |
| | | GS and SCO = 44 overlapping citations. | | | | | | | | | | | | | | | | | | | |
| **KOUSHA & THELWALL& REZAIE (2011)** | Comparisons of citation counts for authored books submitted to 7 social sciences and humanities disciplines in the 2008 (n=1,000) | Citations | 39,733 | | | 12,462 | | | | | | | | | | | | 17,905 | | | |
| | | Google Books and Google Scholar citations were 143 and 18% of Scopus citations, respectively. | | | | | | | | | | | | | | | | | | | |
| **KOUSHA & THELWALL & REZAIE (2011)** | Relative overlap and unique citations for GS and Scopus for 100 sampled authored books submitted to the 2008 Research Assessment Exercise | Citations | 2,599 | | | 789 | | | | | | | | | | | | | | | |
| | | Overlapping GS and SCO = 431 | | | | | | | | | | | | | | | | | | | |
| | | Relative overlap | 16.58 | | | 54.63 | | | | | | | | | | | | | | | |
| | | Unique citations | 2,168 | | | 358 | | | | | | | | | | | | | | | |
| | | (%) | 83.42 | | | 45.37 | | | | | | | | | | | | | | | |
| **BAR-ILAN (2010)** | Citations to "Introduction to informetrics" book | Unique citations | 109 | 46 | | 8 | | | | | | | | | | | | | | | |
| | | (%) | 27.46 | 11.59 | | 2.20 | | | | | | | | | | | | | | | |
| | | GS and WoS= 24 (6.05 %); GS and SCO= 21 (5.29%); SCO and WoS=36 (9.7%); GS and SCO and WoS = 153 (38.54%) | | | | | | | | | | | | | | | | | | | |
| **JAĆIMOVIĆ & PETROVIĆ & ŽIVKOVIĆ (2010)** | Number of cited articles from SDJ | Unique citations | 144 | 4 | | 9 | | | | | | | | | | | | | | | |
| | | (%) | 57.84 | 1.61 | | 3.61 | | | | | | | | | | | | | | | |
| | | GS and WoS= 2 (0.80 %); GS and SCO = 6 (2.41%); SCO and WoS = 37 (14.86%); GS and SCO and WoS = 37 (14.86%) | | | | | | | | | | | | | | | | | | | |
| **MINGERS & LIPITAKIS (2010)** | Publications from 3 UK Business Schools. (n=4,600) | Publications | 3,023 | 1,004 | | | | | | | | | | | | | | | | | |
| | | (%) | 65.72 | 21.83 | | | | | | | | | | | | | | | | | |
| **ŠEMBER & UTROBIČIĆ & PETRAK (2010)** | Croatian medical journal indexed articles (2005-2006) | Unique citations | 86 | 12 | | 39 | | | | | | | | | | | | | | | |
| | | (%) | 22 | 3 | | 10 | | | | | | | | | | | | | | | |
| | | GS and WoS= 9 (2 %); GS and SCO = 36 (9 %); SCO and WoS = 47 (12%); GS and SCO and WoS = 166 (42%) | | | | | | | | | | | | | | | | | | | |
| **BORNMANN et al (2009)** | Papers published (n=1,837) by the journal AngewandteChemie | Publications | 1,747 | | | 1,827 | | | | | | | | | | 1,837 | 1,837 | | | | |
| | | (%) | 95.1 | | | 99.5 | | | | | | | | | | 100 | 100 | | | | |
| | | Citations | 9,320 | | | 44,601 | | | | | | | | | | 44,502 | 48,160 | | | | |
| **FRANCESCHET (2009)** | Publications and cites of computer scientist group | Publications | 1,776 | 324 | | | | | | | | | | | | | | | | | |
| | | (%) | 84.57 | 15.43 | | | | | | | | | | | | | | | | | |
| | | Citations | 10,690 | 1,378 | | | | | | | | | | | | | | | | | |
| **JACOBS (2009)** | Comparison of citation counts for 30 Top articles in Gender & Society | Citations | 8,047 | 3,667 | | | | | | | | | | | | | | | | | |
| | | Ratio of Google to ISI = 2.19 | | | | | | | | | | | | | | | | | | | |
| **KULKARNI et al (2009)** | Cohort study of 328 articles published | Citations | 83,538 | 68,088 | | 82,076 | | | | | | | | | | | | | | | |
| **MARTELL (2009)** | Title search, citations and average citations per article (n=217) | Citations | 1,394 | 680 | | | | | | | | | | | | | | | | | |
| | | Average citations per article | 6.4 | 3.1 | | | | | | | | | | | | | | | | | |
| **MIKKI (2009)** | GS is compared with WoS for earth science authors (n=29) | Publications | 5,048 | 1,573 | | | | | | | | | | | | | | | | | |
| | | (%) | 76.25 | 23.76 | | | | | | | | | | | | | | | | | |
| | | Citations | 40,908 | 43,028 | | | | | | | | | | | | | | | | | |
| | | Average h-index | 16.0 | 16.7 | | | | | | | | | | | | | | | | | |
| **MOSKOVKIN (2009)** | Publications of the 10 largest universities (2008) | Publications | 565,709 | 55,581 | | | | | | | | | | | | | | | | | |
| | | (%) | 91.06 | 8.94 | | | | | | | | | | | | | | | | | |
| **ONYANCHA & OCHOLLA & (2009)** | Comparison of 10 purposefully selected LIS researchers in South Africa | Publications | 384 | 182 | | 96 | | | | | | | | | | | | | | | |
| | | (%) | 58 | 27.50 | | 14.50 | | | | | | | | | | | | | | | |
| | | Citations | 887 | 125 | | 190 | | | | | | | | | | | | | | | |
| | | Average H-index | 5 | 1.7 | | 2.3 | | | | | | | | | | | | | | | |
| **HARZING & VAN DER WAL (2008)** | Comparison between WoS and GS for the impact of books between 1991-2001 | Citations | 883 | 346 | | | | | | | | | | | | | | | | | |
| | | GS reports 2.5 times as many citations as WoS. | | | | | | | | | | | | | | | | | | | |
| **KOUSHA & THELWALL (2008)** | A sample of 882 articles from 39 open access ISI-indexed journals in 2001 | Citations | 5,589 | 4,184 | | | | | | | | | | | | | | | | | |
| | | Unique citations | 3,202 | 1,797 | | | | | | | | | | | | | | | | | |
| | | ISI citations overlapping with Google Scholar = 2,387 | | | | | | | | | | | | | | | | | | | |
| **BAR-ILAN (2007)** | Compares the h-index of highly-cited Israeli researchers | Average H-index | 17.55 | 17.3 | | 17.1 | | | | | | | | | | | | | | | |
| | | Average citations | 245.64 | 162.85 | | 170.27 | | | | | | | | | | | | | | | |
| **MEHO & YANG (2007)** | Citations to 25 library and information science (LIS) faculty members (n=5,285) | Distribution of unique and overlapping citations | GS =2,552 (48.3%); SCO AND WoS=1,104 (20.9%); GS, SCO AND WoS=1,629 (30.8%); GS identifies 1,448 (53.0%) more citations than WoS and Scopus together (4,181 citations for GS in comparison to 2,733 for the union of WoS and Scopus) | | | | | | | | | | | | | | | | | | |
| **BAKKALBASI et al (2006)** | 11 journal titles from each discipline (oncology) using the JCR. All articles (n=614) published 1993- 2003 | Unique Citations | 78 | 41 | | 74 | | | | | | | | | | | | | | | |
| | | (%) | 13 | 7 | | 12 | | | | | | | | | | | | | | | |
| | | GS AND WoS = 26 (4%); GS AND SCO = 31 (5%); WoS AND SCO = 175 (28%); GS, WoS AND SCO = 189 (31%) | | | | | | | | | | | | | | | | | | | |
| **BAKKALBASI et al (2006)** | 11 journal titles (condensed matter physics) using the JCR. All articles(n=296) published 1993-2003 | Unique citations | 50 | 63 | | 25 | | | | | | | | | | | | | | | |
| | | (%) | 17 | 20 | | 8 | | | | | | | | | | | | | | | |
| | | GS AND WoS = 21 (9%); GS AND SCO = 9 (3%); WoS AND SCO = 65 (22%); GS, WoS AND SCO = 63 (21%) | | | | | | | | | | | | | | | | | | | |
| **SALISBURY & TEKAWADE (2006)** | Journal coverage for Agricultural Economics and AgriBusiness (2004-2005) | No. of Titles | 184 | | | | | | | | | | | 92 | 133 | | | | | | |
| | | Average year | 92 | | | | | | | | | | | 46 | 66.5 | | | | | | |
| | | % (n=108) | 85.19 | | | | | | | | | | | 42.59 | 61.57 | | | | | | |
| | | Citations (%) | 39 | | | | | | | | | | | 17 | 44 | | | | | | |
| **YANG & MEHO (2006)** | Items published by two Library and Information Science full-time faculty members. | Unique citations | 38 | 89 | | 25 | | | | | | | | | | | | | | | |
| | | (%) | 9.9 | 23.1 | | 6.5 | | | | | | | | | | | | | | | |
| **JACSO (2005b)** | Citations count for the papers published in 22 volumes of APJAI (n=698) | Documents | 680 | 675 | | | | | | | | | | | | | | | | | |
| | | (%) | 97.42 | 96.60 | | | | | | | | | | | | | | | | | |
| | | Citations | 595 | 1,355 | | | | | | | | | | | | | | | | | |
| **NORUZI (2005)** | Citation counts from Google Scholar and Web of Science (WoS) for Almind & Ingwersen | Citations | 98 | 81 | | | | | | | | | | | | | | | | | |
| | | Unique citations | 64 | 47 | | | | | | | | | | | | | | | | | |
| | | Citations GS AND WoS = 34 | | | | | | | | | | | | | | | | | | | |
| **NORUZI (2005)** | Average cites of the most-cited 36 Authors in the field of Webometrics on GS and WoS | Citations | 1,110 | 729 | | | | | | | | | | | | | | | | | |
| | | Average citations | 30.84 | 20.25 | | | | | | | | | | | | | | | | | |

## ACRONYMS:

GS/GSM/GSC: Google Scholar / Google Scholar Metrics/ Google Scholar Citations
WoS: Web of Science
MAS: Microsoft Academic Search
SCO: Scopus
PUB: Pubmed
PSY: PsycINFO
ERIC: Education Resources Information Center
SSCI: Social Sciences Citation Inde
SJR: SCImago Journal Rank
JCR: Journal Citation Reports
ECON: Econlit
CAB: CAB ABSTRACTS
SCI: Science Citation Index
CA: Chemical Abstracts
GB: Google Books
KoMCI: Korean Medical Citation Index
KMS: KoreaMed Synapse
CINAHL: Cumulative Index to Nursing and Allied Health Literature

## APPENDIX II. EMPIRICAL STUDIES ABOUT GOOGLE SCHOLAR ACCORDING TO UNIT OF ANALYSIS AND DOCUMENT TYPE

| AUTHORS | SAMPLE | TYPE ANALYSIS GS | UNIT | GS/GSM/GSC | % | WoS | % | SCO | % |
|---|---|---|---|---|---|---|---|---|---|
| **WINTER, & ZADPOOR & DODOU (2014)** | Number of citations on 5 April 2013 to Garfield (1955) for WoS and GS as a function of document type | Citations | Journals | 805 | 69.6 | 546 | 90.1 | | |
| | | | Conferences | 123 | 10.6 | 53 | 8.7 | | |
| | | | Books or book chapters | 63 | 5.4 | 7 | 1.2 | | |
| | | | Theses | 75 | 6.5 | 0 | 0 | | |
| | | | Reports | 13 | 1.1 | 0 | 0 | | |
| | | | Other | 43 | 3.7 | 0 | 0 | | |
| | | | Unknown | 34 | 2.9 | 0 | 0 | | |
| | | | Duplicates | 64 | - | 0 | 0 | | |
| | | | False positives | 11 | - | 1 | - | | |
| | | | All types (incl. duplicates and false positives) | 1,231 | | 607 | | | |
| **MIRI & RAOOFI & HEIDARI (2012)** | Comparison of Number of Citations in ISI, GS, and SC based on Article Types Published in Hepatitis Monthly (2008, 2009) | Citations | Original Article | 39 | | 30 | | 40 | |
| | | | Review Article | 28 | | 25 | | 25 | |
| | | | Brief Report | 9 | | 8 | | 10 | |
| | | | Editorial | 8 | | 4 | | 7 | |
| | | | Case Report | 3 | | 3 | | 3 | |
| | | | Letter to the Editor | - | | 1 | | 1 | |
| | | | Guidelines and Clinical Algorithm | - | | - | | - | |
| **LASDA BERGMAN (2012)** | Source types of citing references | Citations | Article | 1,951 | 59.6 | 1,735 | 99.7 | 1,782 | 83.8 |
| | | | Book | 318 | 9.7 | - | - | 1 | 0.0 |
| | | | Conference Paper | 32 | 1.0 | - | - | 25 | 1.2 |
| | | | Foreign Language | 281 | 8.6 | - | - | - | - |
| | | | Government Document | 44 | 1.3 | - | - | - | - |
| | | | Dissertation | 329 | 10.1 | - | - | - | - |
| | | | Master's Thesis | 108 | 3.3 | - | - | - | - |
| | | | Bachelor's Thesis | 6 | 0.2 | - | - | - | - |
| | | | Report | 84 | 2.6 | - | - | - | - |
| | | | Syllabus | 5 | 0.2 | - | - | - | - |
| | | | Unpublished Manuscript | 44 | 1.3 | - | - | - | - |
| | | | Working Paper | 35 | 1.1 | - | - | - | - |
| | | | Review | 24 | 0.7 | - | - | 248 | 11.7 |
| | | | Presentation Slides | 3 | 0.1 | - | - | - | - |
| | | | Blog | 3 | 0.1 | - | - | - | - |
| | | | Editorial | 1 | 0.0 | - | - | 28 | 1.3 |
| | | | Letters to the Editor | 1 | 0.0 | - | - | 10 | 0.5 |
| | | | Supplementary Material | 1 | 0.0 | - | - | - | - |
| | | | Web Page | 1 | 0.0 | - | - | - | - |
| | | | Guideline | 1 | 0.0 | - | - | - | - |
| | | | Series | - | - | 6 | 0.3 | - | - |
| | | | Short Survey | - | - | - | - | 5 | 0.2 |
| | | | Note | - | - | - | - | 27 | 1.3 |
| | | | Total | 3,272 | 100.0 | 1,741 | 100.0 | 2,126 | 100.0 |
| **MEHO & YANG (2007)** | Citations to the work of 25 LIS faculty members Citation count by document type (1996 –2005). | Citations | Journal articles | 2,215 | 40.32 | 1,529 | 75.6 | 1,754 | 76.2 |
| | | | Conference papers | 1,849 | 33.66 | 229 | 11.3 | 359 | 15.6 |
| | | | Review articles | 86 | 1.57 | 172 | 8.5 | 147 | 6.4 |
| | | | Editorial materials | 25 | 0.46 | 63 | 3.1 | 36 | 1.6 |
| | | | Book reviews | 3 | 0.05 | 17 | 0.8 | 0 | 0.0 |
| | | | Letters to the editor | 2 | 0.04 | 9 | 0.4 | 2 | 0.1 |
| | | | Biographical item | 1 | 0.02 | 2 | 0.1 | 1 | 0.0 |
| | | | Doctoral dissertations | 261 | 4.75 | - | - | - | - |
| | | | Master's theses | 243 | 4.42 | - | - | - | - |
| | | | Book chapters | 199 | 3.62 | - | - | - | - |
| | | | Technical reports | 129 | 2.35 | - | - | - | - |
| | | | Reports | 110 | 2.00 | - | - | - | - |
| | | | Books | 102 | 1.86 | - | - | - | - |
| | | | Conference presentations | 72 | 1.31 | - | - | - | - |
| | | | Unpublished papers | 65 | 1.18 | - | - | - | - |
| | | | Bachelor's theses | 34 | 0.62 | - | - | - | - |
| | | | Working papers | 31 | 0.56 | - | - | - | - |
| | | | Research reports | 23 | 0.42 | - | - | - | - |
| | | | Workshop papers | 15 | 0.27 | - | - | - | - |
| | | | Doctoral dissertation proposals | 9 | 0.16 | - | - | - | - |
| | | | Conference posters | 9 | 0.16 | - | - | - | - |
| | | | Book reviews | 3 | 0.05 | - | - | - | - |
| | | | Master's thesis proposals | 3 | 0.05 | - | - | - | - |
| | | | Preprints | 3 | 0.05 | - | - | - | - |
| | | | Conference paper proposals | 2 | 0.04 | - | - | - | - |
| | | | Government documents | 2 | 0.04 | - | - | - | - |
| | | | Total | 5,493 | 100.00 | 2,023 | 100.0 | 2,301 | 100.0 |
| | | | Total from journals | 2,332 | 42.45 | 1,794 | 88.7 | 1,942 | 84.4 |
| | | | Total from conference papers | 1,849 | 33.66 | 229 | 11.3 | 359 | 15.6 |
| | | | Total from journals and conferences | 4,181 | 76.12 | 2,023 | 100.0 | 2,301 | 100.0 |
| | | | Total from dissertations/theses | 538 | 9.79 | - | - | - | - |
| | | | Total from books | 301 | 5.48 | - | - | - | - |
| | | | Total from reports | 262 | 4.77 | - | - | - | - |
| | | | Total from other document types | 211 | 3.84 | - | - | - | - |
| **YANG & MEHO (2006)** | Breakdown of Citations Found in Google Scholay by Document Type by two Library and Information Science full-time faculty members. In | Citations | Journal Articles | 169 | 48,4 | | | | |
| | | | Conference Papers | 90 | 25,8 | | | | |
| | | | Research reports | 39 | 11,2 | | | | |
| | | | Dissertations and Theses | 15 | 4,3 | | | | |
| | | | Dead links | 7 | 2,0 | | | | |
| | | | Editorial Materials No access | 6 | 1,7 | | | | |
| | | | Workshops | 5 | 1,4 | | | | |
| | | | No access | 4 | 1,1 | | | | |
| | | | Technical reports | 3 | 0,9 | | | | |
| | | | Websites | 3 | 0,9 | | | | |
| | | | Other (chapters, bibliographies) | 8 | 2,3 | | | | |
| | | | Total | 349 | 100 | | | | |

| AUTHORS | SAMPLE | TYPE ANALYSIS GS | PUBLICATION TYPES | N | % OF OUTPUTS | NO. OF PUBS FOUND IN GS | NO. OF PUBS FOUND IN WOS | % GS | % WOK | NO. OF CITATIONS FOUND IN GS | NO. OF CITATIONS FOUND IN WOS | GS CITATION PER PAPER (CPP) | WOS CITATION PER PAPER (CPP) |
|---|---|---|---|---|---|---|---|---|---|---|---|---|---|
| **MINGERS & LIPITAKIS (2010)** | GS and WoS citations by publication type.<br><br>No. Publications from 3 UK Business Schools. (n= 4,600) | Cites/Documents | Total books | 95 | 2.1 | 70 | | | | 2,257 | | 32.24 | |
| | | | Books A | 45 | 2.3 | 38 | | 84.4 | | 1,285 | | 33.8 | |
| | | | Books B | 31 | 2.1 | 21 | | 67.7 | | 567 | | 27.0 | |
| | | | Books C | 19 | 1.6 | 11 | | 57.9 | | 405 | | 36.8 | |
| | | | Total edited books | 76 | 1.7 | 58 | | | | 1,763 | | 30.40 | |
| | | | Edited Books A | 48 | 2.5 | 39 | | 81.3 | | 1,394 | | 35.7 | |
| | | | Edited Books B | 16 | 1.1 | 11 | | 68.8 | | 56 | | 5.1 | |
| | | | Edited Books C | 12 | 1.0 | 8 | | 66.7 | | 313 | | 39.1 | |
| | | | Total book chapters | 619 | 13.4 | 287 | | | | 1,946 | | 6.78 | |
| | | | Book Chapters A | 326 | 16.9 | 149 | | 45.7 | | 1,178 | | 7.9 | |
| | | | Book Chapters B | 184 | 12.6 | 74 | | 40.2 | | 289 | | 3.9 | |
| | | | Book Chapters C | 109 | 9.0 | 64 | | 58.7 | | 479 | | 7.5 | |
| | | | Total journal articles | 2,109 | 45.8 | 1,882 | 1,004 | | | 27,606 | 8,434 | 14.67 | 8.40 |
| | | | Journal Articles A | 801 | 41.4 | 705 | 403 | 88.0 | 50.3 | 15,167 | 4,554 | 21.5 | 11.3 |
| | | | Journal Articles B | 715 | 49.1 | 629 | 309 | 88.0 | 43.2 | 6,831 | 2,361 | 10.9 | 7.6 |
| | | | Journal Articles C | 593 | 48.9 | 548 | 292 | 92.4 | 49.2 | 5,608 | 1,519 | 10.2 | 5.2 |
| | | | Total conference papers | 1,013 | 22.0 | 340 | | | | 848 | | 2.49 | |
| | | | Conference Papers A | 298 | 15.4 | 73 | | 24.5 | | 151 | | 2.1 | |
| | | | Conference Papers B | 356 | 24.5 | 99 | | 27.8 | | 240 | | 2.4 | |
| | | | Conference Papers C | 359 | 29.6 | 168 | | 46.8 | | 457 | | 2.7 | |
| | | | Total working papers | 417 | 8.8 | 286 | | | | 1,535 | | 5.37 | |
| | | | Working Papers A | 317 | 16.4 | 235 | | 74.1 | | 1340 | | 5.7 | |
| | | | Working Papers B | 5 | 0.3 | 1 | | 20.0 | | 0 | | 0.0 | |
| | | | Working Papers C | 85 | 7.0 | 50 | | 58.8 | | 195 | | 3.9 | |
| | | | Total reports | 171 | 3.7 | 59 | | | | 491 | | 8.32 | |
| | | | Reports A | 79 | 4.1 | 32 | | 40.5 | | 306 | | 9.6 | |
| | | | Reports B | 62 | 4.3 | 14 | | 22.6 | | 61 | | 4.4 | |
| | | | Reports C | 30 | 2.5 | 13 | | 43.3 | | 124 | | 9.5 | |
| | | | Total others | 110 | 2.4 | 41 | | | | 133 | | 3.24 | |
| | | | Others A | 19 | 1.0 | 10 | | 52.6 | | 77 | | 7.7 | |
| | | | Others B | 86 | 5.9 | 27 | | 31.4 | | 37 | | 1.4 | |
| | | | Others C | 5 | 0.4 | 4 | | 80.0 | | 19 | | 4.8 | |
| | | | Total | 4,600 | | 3,023 | 1,004 | | | 36,579 | 8,434 | 12.1 | 8.4 |
| | | | Total A | 1,933 | 100.0 | 1,281 | 403 | 66.3 | 50.3 | 20,898 | 4,554 | 16.3 | 11.3 |
| | | | Total B | 1,455 | 100.0 | 876 | 309 | 60.2 | 43.2 | 8,081 | 2,361 | 9.2 | 7.6 |
| | | | Total C | 1,212 | 100.0 | 866 | 292 | 71.5 | 49.2 | 7,6 | 1,519 | 8.8 | 5.2 |

| AUTHORS | SAMPLE | TYPE ANALYSIS GS | UNIT | GS/GSM/GSC | WoS | SCO |
|---|---|---|---|---|---|---|
| **JAĆIMOVIĆ & PETROVIĆ & ŽIVKOVIĆ (2010)** | SDJ citation was collected in September 2010 | Cites | Article | 117 | 50 | 56 |
| | | | Review | 13 | 4 | 6 |
| | | | Editorial | 3 | 1 | 3 |
| | | | Proceedings | 3 | 2 | 3 |
| | | | Miscellaneous | 8 | - | - |
| **BAKKALBASI et al (2006)** | 11 journal titles from each discipline (1993-2003). | Cites | | Oncology | CM Phys | |
| | | | Journal | 31 (62 %) | 18 (37%) | |
| | | | Archive | 3 (6%) | 12 (25%) | |
| | | | College or University | 9 (18%) | 6 (13%) | |
| | | | Government | 3 (6%) | 4 (8%) | |
| | | | Non-Governmental Organization | 2 (4%) | 8 (17 %) | |
| | | | Commercial | 0 | 0 | |
| | | | Other | 2 (4%) | 0 | |
| | | | Total | 50 | 48 | |

| AUTHORS | SAMPLE | TYPE ANALYSIS GS | PUBLICATION TYPES | GS | | WoS | | SCOPUS | | TOTAL | |
|---|---|---|---|---|---|---|---|---|---|---|---|
| | | | | NUMBER OF CITED | NUMBER OF RECEIVED CITATIONS | NUMBER OF CITED | NUMBER OF RECEIVED CITATIONS | NUMBER OF CITED | NUMBER OF RECEIVED CITATIONS | NUMBER OF CITED | NUMBER OF RECEIVED CITATIONS |
| **JAĆIMOVIĆ & PETROVIĆ & ŽIVKOVIĆ (2010)** | Type of cited articles from SDJ was collected in September 2010 | Cites | Informative article | 23 | 43 | 13 | 14 | 18 | 22 | 32 | 55 |
| | | | Original scientific article | 57 | 86 | 39 | 50 | 38 | 50 | 76 | 119 |
| | | | Case repot | 5 | 5 | 2 | 3 | 2 | 3 | 5 | 6 |
| | | | Proceedings | 20 | 31 | 7 | 7 | 5 | 5 | 26 | 37 |
| | | | Review | 7 | 16 | 4 | 6 | 4 | 6 | 8 | 17 |
| | | | Proffesional article | 3 | 6 | 4 | 5 | 5 | 7 | 8 | 12 |
| | | | Preliminary communication | 1 | 1 | 0 | 0 | 0 | 0 | 1 | 1 |
| | | | Article from praxis | 0 | 0 | 0 | 0 | 1 | 1 | 1 | 1 |
| | | | Book review | 1 | 1 | 0 | 0 | 0 | 0 | 1 | 1 |
| | | | Total | 117 | 189 | 69 | 85 | 73 | 94 | 158 | 249 |

| AUTHORS | SAMPLE | TYPE ANALYSIS GS | UNIT | GS/GSM/GSC | % | WoS | % | SCO | % | CA | % |
|---|---|---|---|---|---|---|---|---|---|---|---|
| **BAR-ILAN (2010)** | Document types of the unique items retrieved by GS (=109) were collected 2008. | Documents | Journal | 28 | 25.7% | | | | | | |
| | | | Proceedings | 25 | 22.9% | | | | | | |
| | | | Thesis | 15 | 13.8% | | | | | | |
| | | | Book chapter | 13 | 11.9% | | | | | | |
| | | | Report | 10 | 9.2% | | | | | | |
| | | | Manuscript | 7 | 6.4% | | | | | | |
| | | | In Chinese | 4 | 3.7% | | | | | | |
| | | | Book | 3 | 2.8% | | | | | | |
| | | | Newsletter | 2 | 1.8% | | | | | | |
| | | | Encyclopedia entry | 1 | 0.9% | | | | | | |
| **LEVINE-CLARK & KRAUS (2007)** | Compare GS and CA for finding chemistry information in six different searches (n=702) | Documents | Journal Articles (n=564) | 482 | 85.5 | | | | | 521 | 92.4 |
| | | | Patent (n=54) | 4 | 7.4 | | | | | 24 | 100 |
| | | | Problem (n=26) | 26 | 100 | | | | | 0 | 0.0 |
| | | | Conference proceedings(n=23) | 11 | 47.8 | | | | | 12 | 52.2 |
| | | | Book(n=21) | 21 | 100 | | | | | 7 | 33.3 |
| | | | Dissertation(n=9) | 9 | 100 | | | | | 5 | 55.6 |
| | | | Other (n= 5) | 5 | 100 | | | | | 2 | 40 |

## APPENDIX III. EMPIRICAL STUDIES ABOUT GOOGLE SCHOLAR ACCORDING TO LANGUAGES

| AUTHORS | SAMPLE | UNIT GS | LANGUAGE | GS/GSM/GSC | % | WoS | % | SCO | % |
|---|---|---|---|---|---|---|---|---|---|
| **DELGADO LÓPEZ-CÓZAR & REPISO (2013)** | Sample of journals from the field of communication studies indexed in three databases (n=277). | Journals | English | 181 | 65.3 | 93 | 87.8 | 153 | 91.6 |
| | | | Spanish | 42 | 15.2 | 6 | 5.7 | 10 | 6 |
| | | | Chinese | 27 | 9.7 | 0 | 0.0 | 1 | 0.6 |
| | | | Portuguese | 24 | 8.7 | 0 | 0.0 | 3 | 1.8 |
| | | | French | 12 | 4.3 | 4 | 3.8 | 4 | 2.4 |
| | | | German | 7 | 2.5 | 0 | 0.0 | 0 | 0.0 |
| | | | Italian | 2 | 0.7 | 1 | 0.9 | 1 | 0.6 |
| | | | Russian | 1 | 0.4 | 0 | 0.0 | 0 | 0.0 |
| | | | Danish | 1 | 0.4 | 0 | 0.0 | 1 | 0.6 |
| | | | Japanese | 3 | 1.1 | 0 | 0.0 | 0 | 0.0 |
| | | | Romanian | 1 | 0.4 | 0 | 0.0 | 0 | 0.0 |
| | | | Polish | 0 | 0.0 | 0 | 0.0 | 0 | 0.0 |
| | | | Croatian | 1 | 0.4 | 1 | 0.9 | 0 | 0.0 |
| | | | Dutch | 0 | 0.0 | 1 | 0.9 | 0 | 0.0 |
| | | | Nowegian | 0 | 0.0 | 0 | 0.0 | 0 | 0.0 |
| **REINA-LEAL & REPISO & DELGADO LÓPEZ-CÓZAR (2013)** | Nursing journals on Google Scholar Metrics | Journals | English | 208 | 61.9 | | | | |
| | | | Chinese | 24 | 7.1 | | | | |
| | | | Portuguese | 12 | 3.6 | | | | |
| | | | Multilangage | 12 | 3,6 | | | | |
| | | | German | 4 | 1.2 | | | | |
| | | | Korean | 15 | 4.5 | | | | |
| | | | Spanish | 19 | 5.7 | | | | |
| | | | French | 11 | 3.3 | | | | |
| | | | Hindi | 1 | 0.3 | | | | |
| | | | Italian | 1 | 0.3 | | | | |
| | | | Persian | 2 | 0.6 | | | | |
| | | | Polish | 1 | 0.3 | | | | |
| | | | Japanese | 19 | 5.7 | | | | |
| | | | Dutch | 3 | 0.9 | | | | |
| | | | Bulgarian | 1 | 0.3 | | | | |
| | | | Catalan | 1 | 0.3 | | | | |
| | | | Italian | 1 | 0.3 | | | | |
| | | | Tukish | 1 | 0.3 | | | | |
| **JAĆIMOVIĆ & PETROVIĆ & ŽIVKOVIĆ (2010)** | Type of cited articles from SDJ was collected in September 2010 | Citations | English | 39 | 27 | 40 | 82.4 | 47 | 69.1 |
| | | | Serbian | 61 | 42.3 | 17 | 29.8 | 17 | 25 |
| | | | Bilingual | 36 | 25 | 0 | 0.0 | 2 | 2.9 |
| | | | Other | 8 | 5.5 | 0 | 0.0 | 2 | 2.9 |
| **KOUSHA & THELWALL (2008)** | A sample of 882 articles from 39 open access ISI-indexed journals in 2001 | Citations | | BIOLOGY % | CHEMISTRY % | PHYSICS % | COMPUTING % | | |
| | | | English | 57 | 65.5 | 96 | 96 | | |
| | | | Chinese | 36 | 23 | 2 | 1.5 | | |
| | | | Non-English | 7 | 11.5 | 2 | 2.5 | | |
| **MEIER & CONKLING ( 2008 )** | Records retrieved from Compendex were searched in Google Scholar, (1950-2007) | Documents | | The range of non-English language content in Compendex varied between 10.8% and 28.8% for the disciplines. The average amount of non-English materials was 20.5% for those years. In this study, only 11.3% of the missed papers in Google Scholar were non-English. | | | | | |
| **MEHO & YANG (2007)** | Citations to the work of 25 LIS faculty members. Citation count distribution by language (1996 –2005) | Citations | English | 3,891 | 93.06 | 2 | 98.86 | 2,285 | 99.30 |
| | | | Portuguese | 92 | 2.20 | | | | |
| | | | Spanish | 63 | 1.51 | 4 | 0.20 | 3 | 0.13 |
| | | | German | 38 | 0.91 | 13 | 0.64 | 9 | 0.39 |
| | | | Chinese | 44 | 1.05 | | | | |
| | | | French | 32 | 0.77 | | | | |
| | | | Italian | 8 | 0.19 | 3 | 0.15 | 1 | 0.04 |
| | | | Japanese | 1 | 0.02 | | | | |
| | | | Swedish | 3 | 0.07 | 3 | 0.15 | 3 | 0.13 |
| | | | Czech | 2 | 0.05 | | | | |
| | | | Dutch | 2 | 0.05 | | | | |
| | | | Finnish | 2 | 0.05 | | | | |
| | | | Croatian | 1 | 0.02 | | | | |
| | | | Hungarian | 1 | 0.02 | | | | |
| | | | Polish | 1 | 0.02 | | | | |
| | | | Non-English | 290 | 6.94 | 23 | 1.14 | 16 | 0.70 |
| | | | Total | 4,181 | 100 | 2,023 | 100 | 2,301 | 100 |
| **NEUHAUS & NEUHAUS & ASHER & WREDE (2006)** | Contents of 47 different databases with that of Google Scholar, (April-July, 2005) | Documents (PsycINFO) | English | | 68 | | | | |
| | | | Non-English | | 12 | | | | |